\documentclass[referee]{aa}
\usepackage{paralist}

\usepackage[]{graphicx}
\usepackage{subfig}
\usepackage{booktabs}
\usepackage{amssymb}
\usepackage{epstopdf}
\usepackage{xcolor}
\usepackage[varg]{txfonts}
\usepackage{lscape}
\usepackage{geometry} 
\usepackage{siunitx}
\usepackage{hyperref}
\usepackage{natbib}
\bibpunct{(}{)}{;}{a}{}{,} 

\def\ergs{\mathrm{erg\,s^{-1}\,cm^{-2}\,\AA^{-1}}}
\def\ergsline{\mathrm{erg\,s^{-1}\,cm^{-2}}}

\newcommand{\lya}{\ensuremath{\mathrm{Ly}\alpha}}

\newcommand{\amind}{\ensuremath{\mathrm{arcmin}^2}}

\newcommand{\ori}{\textsf{ORIGIN}}
\newcommand{\udft}{\textsf{udf-10}}
\newcommand{\mosaic}{\textsf{mosaic}}
\newcommand{\noisechisel}{\textsf{NoiseChisel}}

\newcommand{\bp}{{\bf{p}}}
\newcommand{\bc}{{\bf{c}}}
\newcommand{\bn}{{\bf{n}}}
\newcommand{\br}{{\bf{r}}}
\newcommand{\bs}{{\bf{s}}}
\newcommand{\bd}{{\bf{d}}}
\newcommand{\bt}{{\bf{t}}}
\newcommand{\bb}{{\bf{b}}}
\newcommand{\bV}{{\bf{V}}}
\newcommand{\bSigma}{{\bf{\Sigma}}}

\newcommand{\rafprobnc}{39}
\newcommand{\mkcatupnum}{200}
\newcommand{\mkcatupnsigma}{1}

\newcommand{\rafcompfilt}{\uppercase{f775w}}
\newcommand{\rafcompbincenter}{27.5}
\newcommand{\rafcompbinpm}{0.25}
\newcommand{\rafcompscmultip}{2}
\newcommand{\rafcompsctolerance}{0.1}
\newcommand{\rafcompdiffstd}{0.13}
\newcommand{\rafcomporigstd}{0.14}

\begin{document} 

 \title{The MUSE Hubble Ultra Deep Field Survey:}
 \subtitle{I. Survey description, data reduction and source detection}

   \author{
           Roland Bacon \inst{1}
           \and Simon Conseil  \inst{1}
           \and David Mary  \inst{2}
           \and Jarle Brinchmann \inst{3,11}
           \and Martin Shepherd \inst{1}
           \and Mohammad Akhlaghi  \inst{1}
           \and Peter M. Weilbacher  \inst{4}
           \and Laure Piqueras \inst{1}
           \and Lutz Wisotzki \inst{4}
           \and David Lagattuta  \inst{1}
           \and Benoit Epinat \inst{5,6}
           \and Adrien Guerou \inst{5,7}         
           \and Hanae Inami  \inst{1}           
           \and Sebastiano Cantalupo \inst{8}
           \and Jean Baptiste Courbot \inst{1,9}
           \and Thierry Contini  \inst{5}
           \and Johan Richard  \inst{1}
           \and Michael Maseda \inst{3}
           \and Rychard Bouwens  \inst{3}
           \and Nicolas Bouch\'e  \inst{5}
           \and Wolfram Kollatschny \inst{10}
           \and Joop Schaye \inst{3}
           \and Raffaella Anna Marino \inst{8}
           \and Roser Pello \inst{5}
           \and Christian Herenz \inst{4}
           \and Bruno Guiderdoni  \inst{1} 
           \and Marcella Carollo  \inst{8} 
           }

   \institute{
   Univ Lyon, Univ Lyon1, Ens de Lyon, CNRS, Centre de Recherche Astrophysique de Lyon UMR5574, F-69230, Saint-Genis-Laval, France
   \and Laboratoire Lagrange, CNRS, Universit\'e C\^ote d'Azur, Observatoire de la C\^ote d'Azur, CS 34229, 06304, Nice, France 
   \and Leiden Observatory, Leiden University, P.O. Box 9513, 2300 RA Leiden, The Netherlands 
   \and Leibniz-Institut f{\"u}r Astrophysik Potsdam,  AIP, An der Sternwarte 16, D-14482 Potsdam, Germany
   \and IRAP, Institut de Recherche en Astrophysique et Plan\'etologie, CNRS,  Universit\'e de Toulouse, 14, avenue Edouard Belin, F-31400 Toulouse, France 
    \and Aix Marseille Universit\'e, CNRS, LAM (Laboratoire d'Astrophysique de Marseille) UMR 7326, 13388, Marseille, France 
    \and ESO, European Southern Observatory, Karl-Schwarzschild Str. 2, 85748 Garching bei Muenchen, Germany
    \and ETH Zurich, Institute of Astronomy, Wolfgang-Pauli-Str. 27, CH-8093 Zurich, Switzerland 
    \and ICube, Universit\'e de Strasbourg - CNRS, 67412 Illkirch, France
   \and Institut  f{\"u}r Astrophysik, Universit{\"a}t G{\"o}ttingen, Friedrich-Hund-Platz 1, D-37077 G{\"o}ttingen, Germany
   \and Instituto de Astrof{\'\i}sica e Ci{\^e}ncias do Espaço, Universidade do Porto, CAUP, Rua das Estrelas, PT4150-762 Porto, Portugal  
   \\
   \\
    \email{roland.bacon@univ-lyon1.fr}
    }

  \date{accepted A\&A 2017-07-25 \\
  }
   
   \thanks{Based on observations made with ESO telescopes at the La Silla Paranal Observatory under programs 094.A-0289(B), 095.A-0010(A), 096.A-0045(A) and 096.A-0045(B)}

  \abstract {
We present the MUSE Hubble Ultra Deep Survey, a mosaic of nine MUSE fields covering 90\% of the entire HUDF region with a 10-hour deep exposure time, plus a deeper 31-hour exposure in a single 1.15 \amind\ field. 
The improved observing strategy and advanced data reduction results in datacubes with sub-arcsecond spatial resolution (0\farcs65 at 7000 \AA) and accurate astrometry (0\farcs07 rms).  
We compare the broadband photometric properties of the datacubes to HST photometry, finding a good agreement in zeropoint up to $m_{AB}=28$ but with an increasing scatter for faint objects. 
 We have investigated the noise properties and developed an empirical way to account for the impact of the correlation introduced by the 3D drizzle interpolation. 
The achieved $3\sigma$ emission line detection limit for a point source is $1.5$ and $3.1\,10^{-19} \ergsline$ for the single ultra-deep datacube and the mosaic, respectively.
We extracted 6288 sources using an optimal extraction scheme that takes the published HST source locations as prior. In parallel, we performed a blind search of emission line galaxies using an original method based on advanced test statistics and filter matching. The blind search results in 1251 emission line galaxy candidates in the mosaic and 306 in the ultradeep datacube, including 72 sources without HST counterparts ($m_{AB}>31$). In addition 88 sources missed in the HST catalog but with clear HST counterparts were identified.
This data set is the deepest spectroscopic survey ever performed. In just over 100 hours of integration time, it provides nearly an order of magnitude more spectroscopic redshifts compared to the data that has been accumulated on the UDF over the past decade. The depth and high quality of these datacubes enables new and detailed studies of the physical properties of the galaxy population and their environments over a large redshift range.
}


   \maketitle
%


\section{Introduction}
\label{sect:intro}

In 2003 the Hubble Space Telescope (HST) performed a 1 Megasecond observation with its Advanced Camera for Surveys (ACS) in a tiny 11 \amind\ region located within the Chandra Deep Field South: the Hubble Ultra Deep Field (HUDF, \citealt{Beckwith+2006}).  The HUDF immediately became the deepest observation of the sky.  This initial observation was augmented a few years later with far ultraviolet images from ACS/SBC \citep{Voyer2009} and with deep near ultraviolet \citep{Teplitz2013} and near infrared imaging \citep{Oesch2010, Bouwens2011, Ellis2013, Koekemoer2013} using the Wide Field Camera 3 (WFC3).  
These datasets  have been assembled into the eXtreme Deep Field (XDF) by \cite{Illingworth2013}. 
With an achieved sensitivity ranging from 29.1 to 30.3 AB mag, this emblematic field is still, fourteen years after the start of the observations, the deepest ever high-resolution image of the sky. Thanks to a large range of ancillary data taken with other telescopes, including for example Chandra \citep{Xue2011, Luo2017}, XMM \citep{Comastri2011}, ALMA \citep{Walter2016, Dunlop2017}, Spitzer/IRAC \citep{Labbe2015}, and the VLA \citep{Kellermann2008, Rujopakarn2016}, the field is also covered at all  wavelengths from X-ray to radio.

Such a unique data set has been central to our knowledge of galaxy formation and evolution at intermediate and high redshifts. For example, \cite{Illingworth2013} have detected 14,140 sources at $5\sigma$ in the field including 7121 galaxies in the deepest (XDF) region.
Thanks to the exceptional panchromatic coverage of the Hubble images (11 filters from 0.3 to 1.6 $\mu m$) it has been possible to derive precise photometric redshifts for a large fraction of the detected sources. In particular, the latest photometric redshift catalog of \citet{Rafelski2015} provides 9927 photometric redshifts up to $z = 8.4$.
This invaluable collection of galaxies has been the subject of many studies spanning a variety of topics, including: the luminosity function of high redshift galaxies (e.g., \citealt{McLure2013, Finkelstein2015, Bouwens2015, Parsa2016}), the evolution of star formation rate with redshift (e.g., \citealt{Ellis2013, Madau2014, Rafelski2016, Bouwens2016, Dunlop2017}), measurements of stellar mass (e.g., \citealt{Gonzalez2011, Grazian2015, Song2016}), galaxy sizes (e.g., \citealt{Oesch2010, Ono2013, vanderWel2014, Shibuya2015, Curtis-Lake2016}) and dust and molecular gas content (e.g., \citealt{Aravena2016a, Aravena2016b, Decarli2016a, Decarli2016b}), along with probes of galaxy formation and evolution along the Hubble sequence (e.g., \citealt{Conselice2011, Szomoru2011}).

Since the release of the HUDF, a significant effort has been made with 8m class ground-based telescopes to perform follow-up spectroscopy of the sources detected in the deep HUDF images. \citet{Rafelski2015} compiled a list of 144 high confidence ground-based spectroscopic redshifts from various instruments and surveys (see their Table 3):
VIMOS-VVDS \citep{Lefevre2004}, FORS1\&2 \citep{Szokoly2004, Mignoli2005} VIMOS-GOODS \citep{Vanzella2005, Vanzella2006, Vanzella2008, Vanzella2009, Popesso2009, Balestra2010} and VIMOS-GMASS \citep{Kurk2013}.
In addition, HST Grism spectroscopy  provided 34 high-confidence spectroscopic redshifts:
GRAPES \citep{Daddi2005} and 3DHST \citep{Morris2015, Momcheva2016}.
This large and long lasting investment in telescope time has thus provided 178 high-confidence redshifts in the HUDF area since 2004.  Although the number of spectroscopic redshifts makes up only a tiny fraction (2\%) of the 9927 photometric redshifts (hereafter photo-z), they are essential for calibrating photo-z accuracy. In particular, by using the reference spectroscopic sample, \citet{Rafelski2015} found that their photo-z measurements achieved a low scatter (less than 0.03 rms in $\mathrm{\sigma_{NMAD}}$) with a reduced outlier fraction (2.4-3.8\%).
 
However, this spectroscopic sample is restricted to bright objects (the median F775W AB magnitude of the sample is 23.7, with only 12\% having AB$>$25) at low redshift: the sample distribution peaks at z $\approx$ 1 and only a few galaxies have z $>$ 2.  The behavior of spectrophotometric methods at high z and faint magnitude is therefore poorly known. Given that most of the HUDF galaxies fall in this regime (96\% of the \citealt{Rafelski2015} sample has AB$>$25 and 55\% has z$>$2), it would be highly desirable to obtain a larger number of high-quality spectra in this magnitude and redshift range. 

Besides calibrating the photo-z sample, though, there are other important reasons to increase the number of sources in the UDF with high quality spectroscopic information.
Some key astrophysical properties of galaxies can only be measured from spectroscopic information, including kinematics of gas and stars, metallicity, and the physical state of gas. Environmental studies also require a higher redshift accuracy than those provided by photo-z estimates. 

The fact that only a small fraction of objects seen in the HST images (representing the tip of the iceberg of the galaxy population) have spectroscopic information shows how difficult these measurements are. In particular, the current state-of-the-art multi-object spectrographs perform well when observing the bright end of galaxy population over wide fields.  But, despite their large multiplex, they are not well adapted to perform deep spectroscopy in very dense environments. An exhaustive study of the UDF galaxy population with these instruments would be prohibitively expensive in telescope time and very inefficient.
Thus, by practical considerations, multi-object spectroscopy is restricted to studying preselected samples of galaxies. Since preselection implies that only objects found in broadband deep imaging will be selected, this technique leaves out potential emission-line only galaxies with faint continua. 

Thankfully, with the advent of MUSE, the Multi Unit Spectroscopic Explorer at the VLT \citep{Bacon+2010} the state of the art is changing. As expressed in the original MUSE science case \citep{Bacon2004}, one of the project's major goals is to push beyond the limits of the present generation of multi-object spectrographs, using the power of integral field spectroscopy to perform deep spectroscopic observations in Hubble deep fields.  

During the last MUSE commissioning run \citep{Bacon+2014} we performed a deep 27-hour integration in a 1 \amind\ region located in the Hubble Deep Field South (hereafter HDFS) to validate MUSE's capability in performing a blind spectroscopic survey. With this data we were able to improve the number of known spectroscopic redshifts in this tiny region by an order of magnitude \citep{HDFS}. This first experiment  not only effectively demonstrated the unique capabilities of MUSE in this context, but has also led to new scientific results: the discovery of extended \lya\ halos in the circumgalactic medium around high redshift galaxies \citep{Wisotzki2016}, the study of gas kinematics  \citep{Contini2016}, the investigation of the faint-end of the \lya\ luminosity function \citep{Drake2016}, the measurement of metallicity gradients \citep{Carton2017} and the properties of galactic winds at high z \citep{Finley2017}.

The HDFS observations also revealed 26 \lya\ emitting galaxies that were not detected in the HST WFPC2 deep broadband images, demonstrating that continuum-selected samples of galaxies, even at the depth of the Hubble deep fields, do not capture the complete galaxy population. This collection of high equivalent width \lya\ emitters found in the HDFS indicates that such galaxies may be an important part of the low-mass, high-redshift galaxy population. However, this first investigation in the HDFS was limited to a small 1 \amind\ field of view and will need to be extended to other deep fields before we can assess its full importance.  

After the HDFS investigation, the next step was to start a more ambitious program on the Hubble Ultra Deep Field. This project was conducted as one of the Guarantee Time Observing (GTO) programs given by ESO in return for the financial investment and staff effort brought by the Consortium to study and build MUSE. This program is part of a wedding cake approach, consisting of the shallower MUSE-Wide survey in the CDFS and COSMOS fields \citep{Wide} covering a wide area, along with a deep and ultra-deep survey in the HUDF field covering a smaller field of view.

This paper (hereafter paper I) is the first paper of a series that describes our investigation of the HUDF and assesses the science results. Paper I focuses on the details of the observations, data reduction, performance assessment and source detection. In paper II \citep{Inami2017} we describe the redshift analysis and provide the source catalog. 
In paper III \citep{Brinchmann2017} we investigate the photometric redshifts properties of the sample. 
The properties of CIII] emitters as \lya\ alternative for redshift confirmation of high-z galaxies are discussed in paper IV \citep{Maseda2017}.
In paper V \citep{Guerou2017} we obtain spatially resolved stellar kinematics of galaxies at z$\approx$0.2-0.8 and compare their kinematical properties with those inferred from gas kinematics. The faint end of the \lya\ luminosity function and its implication for reionisation are presented in paper VI \citep{Drake2017}.
The properties of Fe ii* emission, as tracer of galactic winds in star-forming galaxies is presented in paper VII \citep{Finley2017b}. Extended \lya\ haloes around individual \lya\ emitters are discussed in paper VIII \citep{Leclercq2017}. The first measurement of the evolution of galaxy merger fraction up to z $\approx$ 6 is presented in paper IX \citep{Ventou2017} and a detailed study of \lya\ equivalent widths properties of the \lya\ emitters is discussed in paper X \citep{Hashimoto2017}.

The paper is organized as follows. After the description of the observations (section \ref{sect:obs}), we explain the data reduction process in detail (section~\ref{sect:datared}). The astrometry and broadband photometric performances are discussed in section~\ref{sect:astrophoto}. We then present the achieved spatial and spectral resolution (section~\ref{sect:psf}), including an original method to derive the spatial PSF when there is no point source in the field. Following that, we investigate in section~\ref{sect:noise} the noise properties in detail and derive an estimate of the limiting emission line source detection. Finally, we explain how we perform source detection and describe an original blind search algorithm for emission line objects (section~\ref{sect:detection}). A summary concludes the paper.
\section{Observations}
\label{sect:obs}


The HUDF was observed over eight GTO runs over two years: September, October, November and December 2014,  August, September, October, and December 2015 and February 2016. A total of 137 hours of telescope in dark time and good seeing conditions have been used for this project. This is the equivalent to 116 hours of open shutter time which translates to 85\% efficiency when including the overheads.

\begin{figure}[htbp]
\begin{center}
\includegraphics[width=\columnwidth]{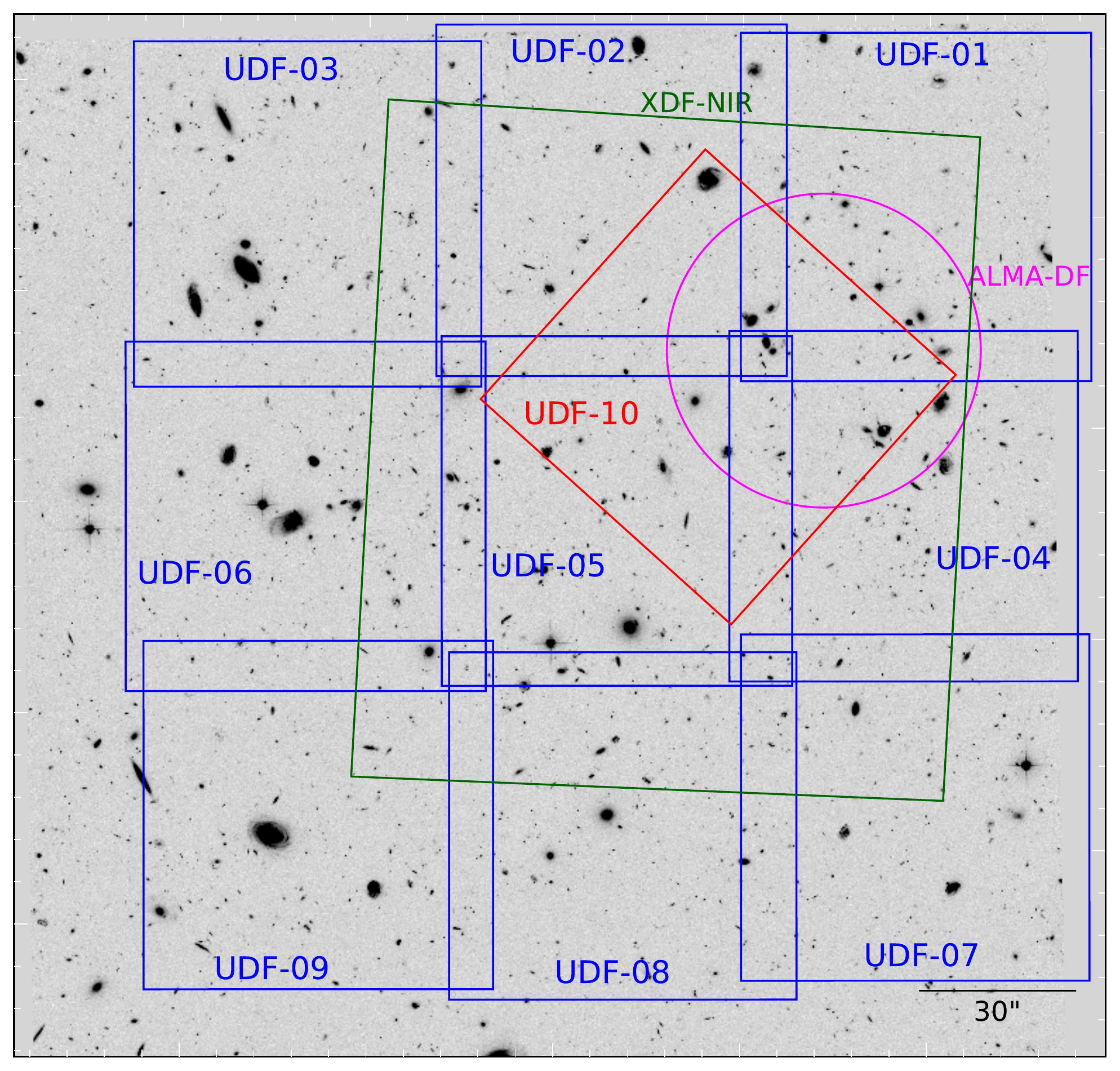}
\caption{Field location and orientation for the \mosaic\ (UDF01-09, in blue) and UDF10 (in red) fields, overlaid on the HST ACS F775W image. The green rectangle indicates the XDF/HUDF09/HUDF12 region containing the deepest near-IR observations from the HST WFC3/IR camera. The magenta circle display the  deep ALMA field from the ASPECS pilot program \citep{Walter2016}. 
North is located 42\degr\ clockwise from the vertical axis.}
\label{fig:fieldslocation}
\end{center}
\end{figure}

\begin{figure}[htbp]
\begin{center}
\includegraphics[width=\columnwidth]{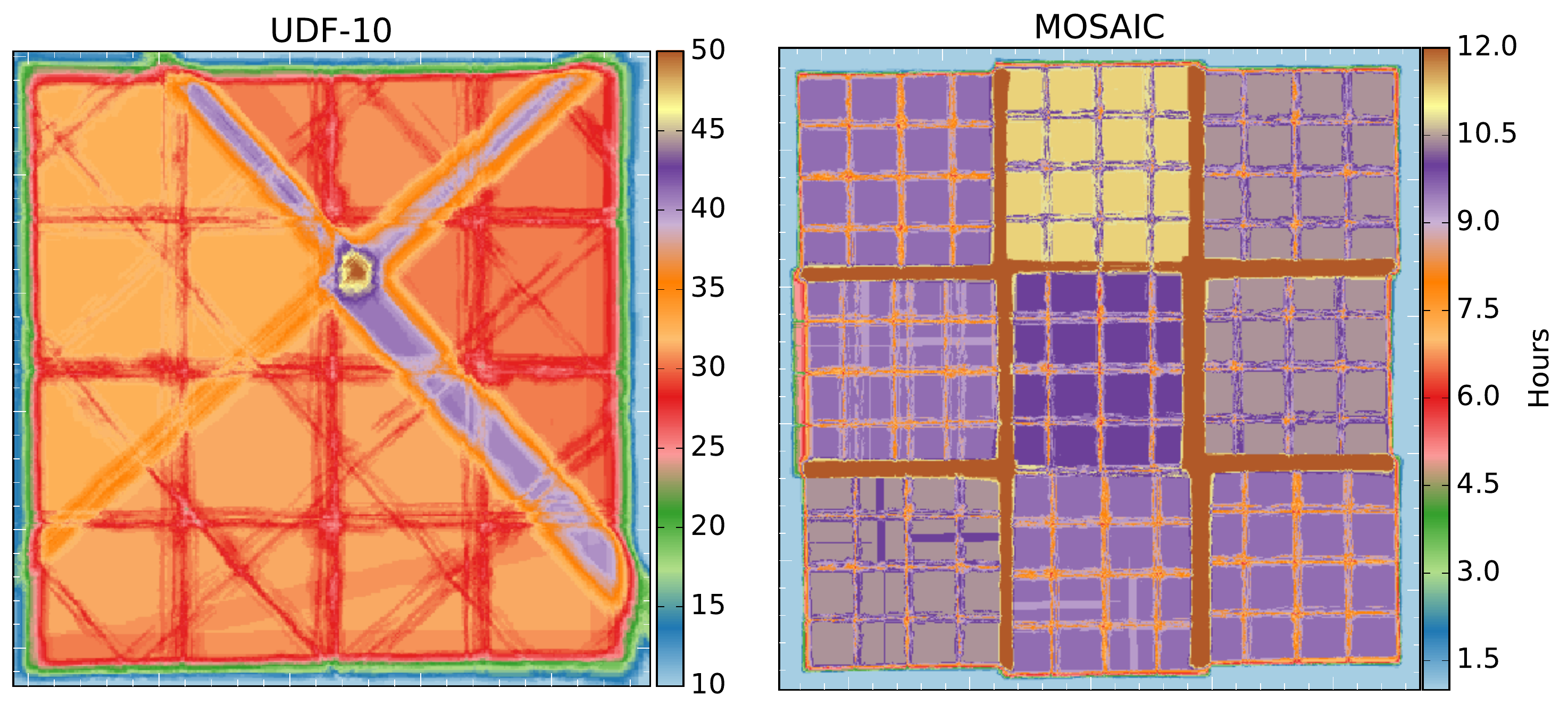}
\caption{Final exposure map images (averaged over the full wavelength range) in hours for the \udft\ and \mosaic\  fields. The  visible stripes correspond to regions of lower integration due to the masking process (see section \ref{sect:mask}).}
\label{fig:expmap}
\end{center}
\end{figure}

\subsection{The medium deep \mosaic\ field}

We covered the HUDF region with a mosaic of nine MUSE fields (UDF-01 through UDF-09, respectively) oriented at a PA of -42\degr\ as shown in Fig~\ref{fig:fieldslocation}.  Each MUSE field is approximately a square $1 \times 1$ \amind\ in area. The dithering pattern used is similar to the HDFS observation scheme \citep{HDFS}: that is, a set of successive 90\degr\ instrument rotations plus random offsets within a 2\arcsec\ square box.

Given its declination (-27\degr\  47\arcmin\ 29\arcsec), the UDF transits very close to zenith in Paranal. When approaching zenith, the rotation speed of the instrument optical derotator increases significantly and its imperfect centering produces a non negligible wobble. However, MUSE has the ability to perform secondary guiding, using stars positioned in a circular ring around the field of view. Image of these stars are affected by the derotator wobble in the same way as the science field, so their shapes can be used to correct for the extra motion. The use of a good slow-guiding star is therefore very important in maintaining field-centering during an exposure, in order to get the best spatial resolution. Thus, the location of each field in the mosaic was optimized to not only provide a small overlap with adjacent fields but also to keep the selected slow-guiding star within the slow-guiding region during the rotation+dither process.  Unfortunately, only a fraction of the fields have an appropriate slow-guiding star within their boundaries (UDF-02, 04, 07, and 08). Therefore, we preferentially observed these fields when the telescope  was near zenith, while the others were observed when the zenith angle was larger than 10\degr.

The integration time for each exposure was 25 minutes. This is long enough to reach the sky-noise-limited regime, even in the blue range of the spectrum, but still short enough to limit the impact of cosmic rays. Including the overheads it is possible to combine two exposures into an observing block spanning approximately 1 hour. A total of 227 25-minute exposures were performed in good seeing conditions. A few exposures were repeated when the requested conditions were not met (e.g., poor seeing or cirrus absorption). As shown in Fig.~\ref{fig:expmap} and taking into account a few more exposures that were discarded for various reasons during the data reduction process (see section \ref{sect:datared}), the \mosaic\ field achieves a depth of $\approx$10 hours over a contiguous area of 9.92 \amind\ within a rectangle approximately $3.15\arcmin \times 3.15\arcmin$\ in shape.

\subsection{The \udft\ ultra deep field}

In addition to the \mosaic, we also performed deeper observations of a single 1\arcmin$\times$1\arcmin\ field,  called UDF-10.  
The field location\footnote{The \udft\ field center is at $\mathrm{\alpha_{J2000}}$ = 03h 32mn 38.7sec, $\mathrm{\delta_{J2000}}$ = -27\degr46\arcmin44\arcsec}  was selected to be in the deepest part of the XDF/HUDF09/HUDF12 area and to overlap as much as possible with the deep ALMA pointing from the ASPECS pilot program \citep{Walter2016}. 
A different PA of 0\degr\ was deliberately chosen to better control the systematics. Specifically, when this field is combined with the overlapping \mosaic\ fields (at a PA of -42\degr), the instrumental slice and channel orientation with respect to the sky is different. This helps to break the symmetry and minimize the small systematics that are left by the data reduction process. Care was taken to have a bright star within the slow-guiding field in order to obtain the best possible spatial resolution, even when the field transits near zenith. Because of this additional constraint, the field only partially overlaps with the deep ALMA pointing. The resulting location is shown in Fig~\ref{fig:fieldslocation}. 

Given that GTO observations are conducted in visitor mode and not in service mode, we performed an equivalent GTO queue scheduling within all GTO observing programs. A fraction of the best seeing conditions were used for this field. During observation, we used the same dithering strategy and individual exposure time as for the \mosaic, obtaining a total of 51 25-minute exposures. 

In the following we call \udft\ the combination of UDF-10 with the overlapping \mosaic\ fields (UDF-01, 02, 04, and 05).
\udft\ covers an area of 1.15 \amind\ and reaches a depth of 31 hours (Fig.~\ref{fig:expmap}). Such a depth is comparable to the 27 hours reached by the HDFS observations \citep{HDFS}. However, as we will see later, the overall quality is much better thanks to the best observing conditions, an improved observational strategy and refined data reduction process.

\section{Data reduction}
\label{sect:datared}

Performing reductions on such a large data set (278 science exposures) is not a negligible task, but the control and minimisation of systematics is extremely important since we want to make optimal use of the depth of the data.  The overall process for the UDF follows the data reduction strategy developed for the HDFS \citep{HDFS} but with improved processes and additional procedures (see \citealt{Conseil2016}). It consists of two major steps: the production of a datacube from each individual exposure and the combination of the datacubes to produce the final \mosaic\ and \udft\ datacubes. These steps are described in the following sections.

\subsection{Data reduction of individual exposures}

\subsubsection{From the raw science data to the first pixtable}
We first run the raw science data through the MUSE standard pipeline version 1.7dev (Weilbacher et al in prep). The individual exposures are processed by the {\em scibasic} recipe which used the corresponding daily calibrations (flatfields, bias, arc lamps, twilight exposures) and geometry table (one per observing run) to produce a table (herafter called {\em pixtable}) containing all pixel information: location, wavelength, photon count and an estimate of the variance. Bad pixels corresponding to known CCD defects (columns or pixels) are also masked at this time. For each exposure we use the {\em illumination} exposure to correct for flux variations at the slices edges due to small temperature changes between the morning calibration exposures and the science exposures. From the adjacent {\em illumination} exposures taken before and after the science, we select the one nearest in temperature.

The pipeline recipe {\em scipost} is then used to perform astrometric and flux calibrations on the {\em pixtable}. We use a single reference flux calibration response for all exposures, created in the following way. All flux calibration responses, obtained over all nights, are scaled to the same mean level to remove transparency variations. Then, we take the median of the stack to produce the final reference response. We note that no sky subtraction is performed at this stage because we use the sky flux to perform self-calibration on each exposure.

A datacube is then created with the {\em makecube} pipeline recipe, using the default 3D drizzling interpolation process. Each exposure needs to be precisely recentered to correct for the derotator wobble. Unlike the HDFS observations, only a few UDF fields have bright point sources that can be used to compute this offset. We have therefore developed an original method to derive precise offset values with respect to the HST reference images. This is described in detail in section~\ref{sect:FSF}. 
The computed ($\Delta \alpha , \Delta \delta$) offset values are then applied to the {\em pixtable}, which is then ready for the self-calibration process.

\subsubsection{Self calibration}
Although the standard pipeline is efficient at removing most of the instrumental signatures, one can still see a low-level footprint of the instrumental slices and channels. This arises from a mix of detector instabilities and imperfect flatfielding, which are difficult to correct for with standard calibration exposures. We therefore use a self-calibration procedure\footnote{The self-calibration procedure is part of the \textsf{MPDAF} software \citep{Piqueras2016}: the MUSE Python Data Analysis Framework. It is an open-source (BSD licensed) Python package, developed and maintained by CRAL and partially funded by the ERC advanced grant 339659-MUSICOS. It is available at https://git-cral.univ-lyon1.fr/MUSE/mpdaf},
similar in spirit to the one used for the HDFS \citep{HDFS} but enhanced to produce a better correction. 
It is also similar to the  \textsf{CubeFIX} flat-fielding correction method, part of the  \textsf{CubExtractor} package developed by Cantalupo (in prep.) and used, for instance, in
\cite{Borisova2016} (see therein for a short description) but it works directly on the {\em pixtable}. Compared to the HDFS version, the major changes in the new procedure are to perform polychromatic correction and to use a more efficient method to reject outliers. 

\begin{figure}[htbp]
\begin{center}
\includegraphics[width=\columnwidth]{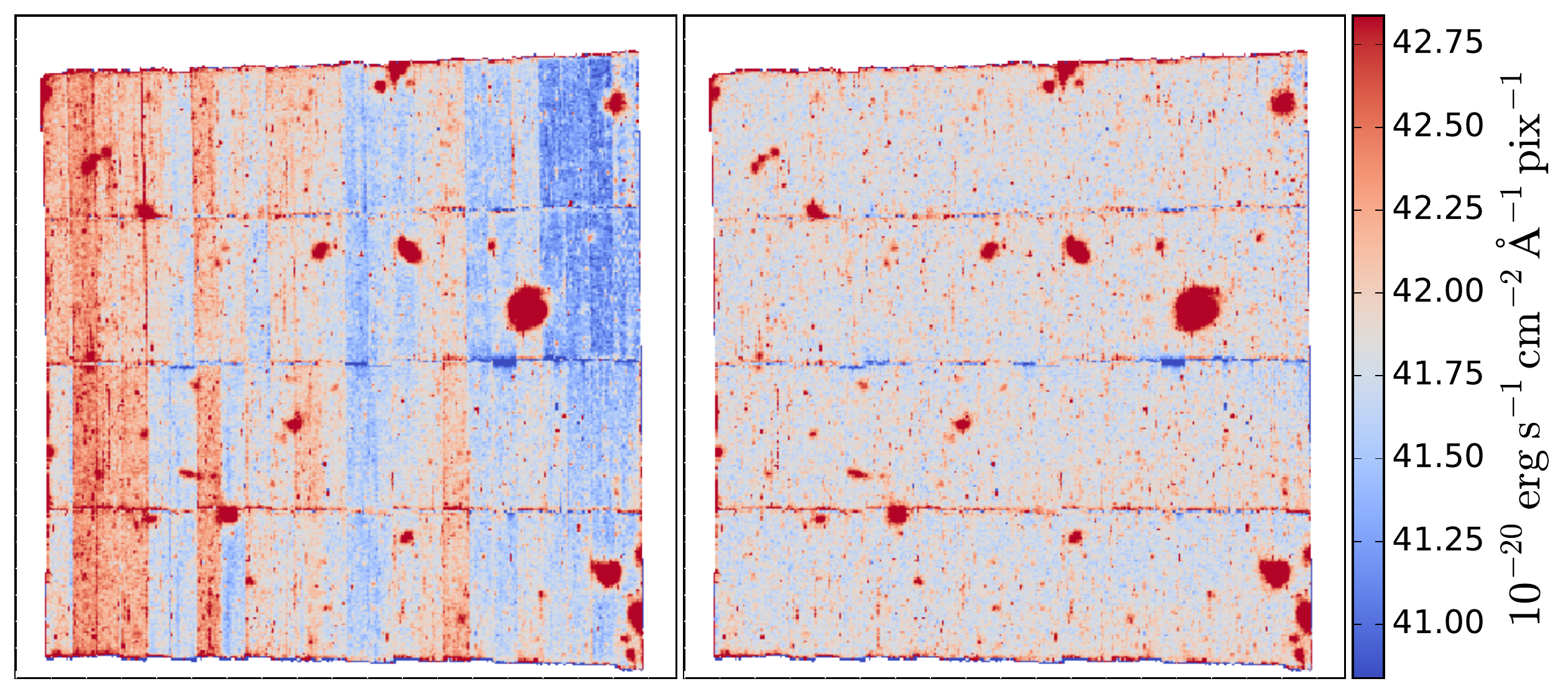}
\caption{Self-calibration on individual exposures. The reconstructed white light image of a single exposure, highly stretched around the mean sky value, is shown before (left panel) and after (right panel) the self calibration process.}
\label{fig:selfcalib}
\end{center}
\end{figure}

The procedure starts by masking all bright objects in the data.  The mask we use is the same for all exposures, calculated from the white light image of the rough, first-pass datacube of the combined UDF data set. The method works on 20 wavelength bins of 200-300 \AA. These bins have been chosen so that their edges do not fall on a sky line. The median flux of each slice\footnote{The slices are the thin mirrors of the MUSE image slicer which perform the reformatting of the entrance field of view into a pseudo slit located at the spectrograph input focal plane} is computed over the wavelength range of the bin, using only the unmasked voxels\footnote{voxel: volume sampling element (0\farcs2 $\times$ 0\farcs2 $\times$ 1.25 \AA)} in the slice.  Individual slices flux are then offset to the mean flux of all slices and channels over the same wavelength bin. Outliers are rejected using 15$\sigma$ clipping based on a the median absolute deviation (MAD). As shown in Fig.~\ref{fig:selfcalib}, the new self calibration is very efficient in removing the remaining flatfielding defects and other calibration systematics.

\subsubsection{Masking}
\label{sect:mask}
Some dark or bright regions at the edges of each slice stack (hereafter called inter-stack defects) can be seen as thin, horizontal strips in Fig.~\ref{fig:selfcalib}. These defects are not corrected by standard flat-fielding or through self-calibration and appear only in deep exposures of the empty field. It is important to mask them because otherwise the combinations of many exposures at different instrumental rotation angles and with various on-sky offsets will impact a broad region on the final data cube.

To derive the optimum mask, we median-combine all exposures, irrespective of the field, projected on an instrumental grid (i.e., we stack based on fixed pixel coordinates instead of the sky's world coordinate system). In such a representation, the instrumental defects are always at the same place, while sky objects move from  place to place according to the dithering process. The resulting mask identifies the precise locations of the various defects on the instrumental grid. This is used to build a specific {\em bad pixel table} which is then added as input to the standard {\em scibasic} pipeline recipe.

In principle, to mask the inter-stack region one can simply produce a datacube using this additional {\em bad pixel table} with the {\em scibasic} and {\em scipost} recipes. However, the 3D drizzle algorithm used in {\em scipost} introduces additional interpolation effects which prevents perfect masking. To improve the inter-stack masking, we run the {\em scibasic} and {\em scipost} recipes twice: the first time without using the specific {\em bad pixel table}, and the second time with it.  Using the output of the ``bad-pixel'' version of the cube, we derive a new, 3D mask which we apply to the original cube, effectively removing the inter-stack bad data. 

Even after this masking, a few exposures had some unique problems which required additional specific masking.  This was the case for 2 exposures impacted by earth satellite trails, and for 9 exposures that show either high dark levels in channel 1 or important bias residuals in channel 6. An individual  mask was built and applied for each of these exposures.  The impact of all masking can be easily seen  in Fig.~\ref{fig:expmap} where the stripes with lower integration time show up in the exposure maps.

\subsubsection{Sky subtraction}
\label{sect:skysub}
The recentered and self-calibrated {\em pixtable} of each exposure is then sky subtracted, using the {\em scipost} pipeline recipe with sky subtraction enabled, and a new datacube is created on a fixed grid. For the \mosaic\ field, we pre-define a single world coordinate system (with a PA of -42\degr) covering the full mosaic region, and each of the nine MUSE fields (UDF-1 through 9) is projected onto the grid. For the \udft\ a different grid is used (PA=0\degr). Based on the overlap region, fields UDF-1, 2, 4, 5 and 10 are projected onto this grid.

\begin{figure}[htbp]
\begin{center}
\includegraphics[width=\columnwidth]{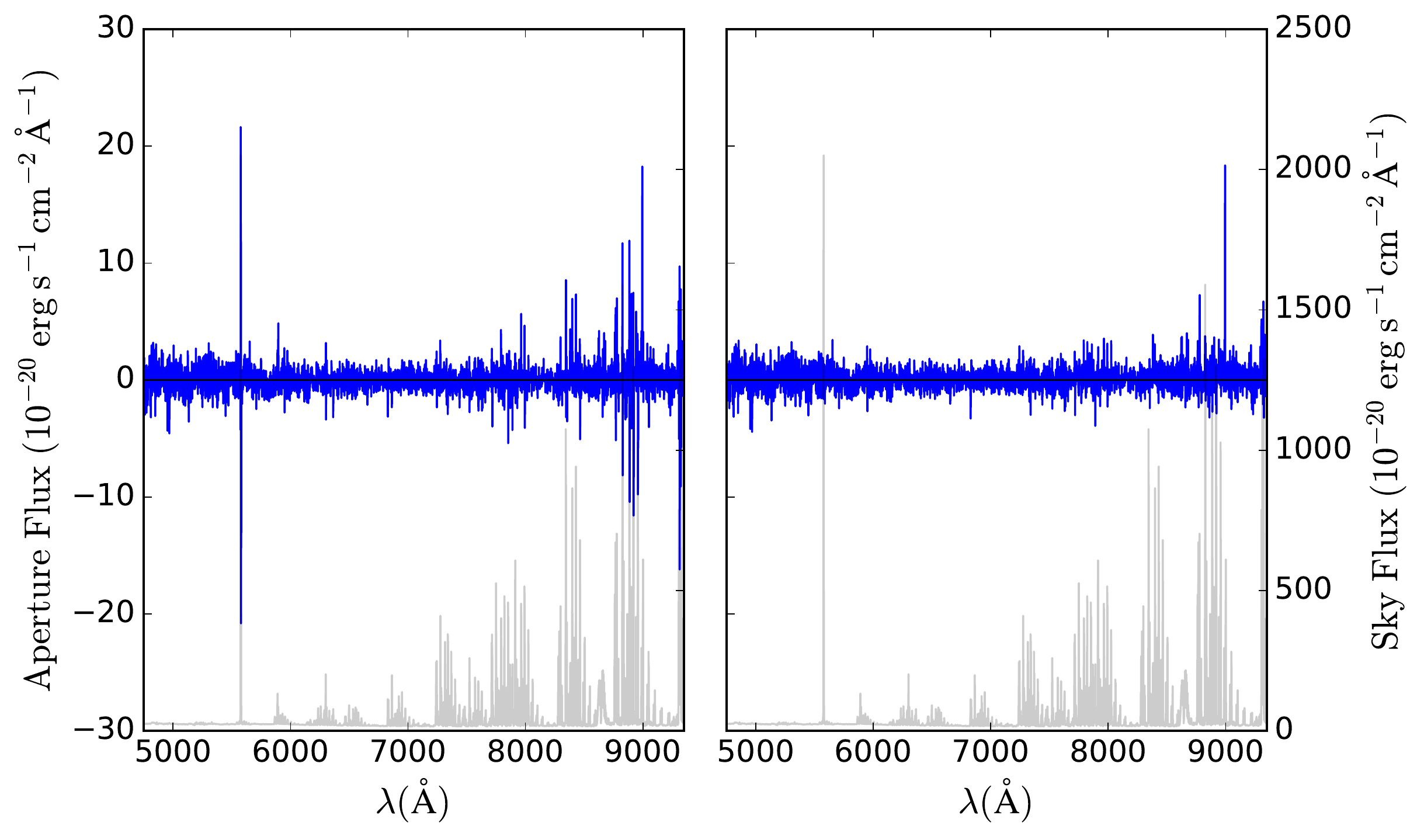}
\caption{Spectrum extracted from a 1\arcsec\ diameter aperture in an empty region of a single exposure datacube, before (left panel) and after (right panel) the use of ZAP. The mean sky spectrum is shown in light gray.}
\label{fig:zap}
\end{center}
\end{figure}

We then used ZAP \citep{ZAP}, the principal component analysis enhanced sky subtraction software developed for MUSE datacubes.  As shown in Fig.~\ref{fig:zap}, ZAP is very efficient at removing the residuals left over by the standard pipeline sky subtraction recipe. The computed inter-stack 3D mask is then applied to the resulting datacube.

\subsubsection{Variance estimation}
\label{sect:variance}
Variance estimation is a critical step that is used to evaluate the achieved signal-to-noise ratio and to perform source extraction and detection, as we will see later in section~\ref{sect:detection}. The pipeline first records an estimate of the variance at each voxel location, using the measured photon counts as a proxy for the photon noise variance and adding the read-out detector variance. This variance estimate is then propagated accurately along each step of the reduction, taking into account the various linear transformations that are performed on the {\em pixtable}.  However, even after accounting for these effects, there are still problems with the variance estimates.

\begin{figure}[htbp]
\begin{center}
\includegraphics[width=\columnwidth]{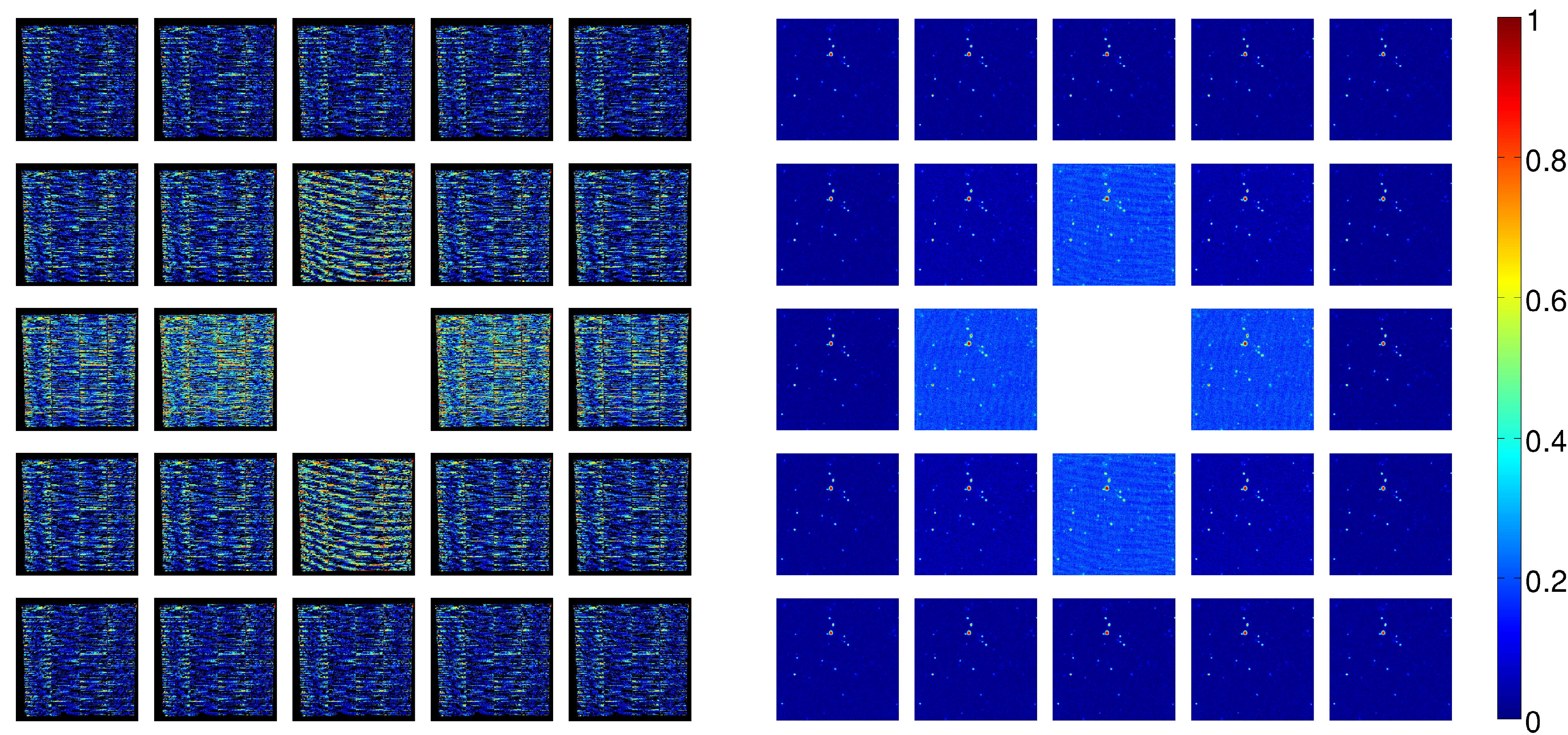}
\caption{Spatially correlated properties in the MUSE \udft\ data cube after drizzle interpolation. Each image shows the correlation between spectra and their $\pm$1, $\pm$2 spatial neighbors. The correlation image is shown for a single exposure datacube (left panel) and for the combined datacube (right panel). Note that the correlation was performed on the blue part of the spectrum to avoid the OH lines region.}
\label{fig:covariance}
\end{center}
\end{figure}

The first problem is that the estimate is noisy, given that the random fluctuations around the unknown mean value are used in place of the mean itself for each pixel. The second problem is related to the interpolation used to build the regular grid of the datacube from the non-regular  {\em pixtable} voxels. This interpolation creates correlated noise in the output datacube as can be seen in Fig.~\ref{fig:covariance}. To take into account this correlation, one should in principle propagate both the variance information and the covariance matrix, instead of just the variance as the pipeline does. However, this covariance matrix is far too large ($\approx$125 times the datacube size, even if we limit it to pixels within the seeing envelope and 5 pixels along the spectral axis) and thus cannot be used in practice.

The consequence is that the pipeline-propagated variance for a single exposure exhibits strong oscillations along both the spatial and spectral axes. When combining multiple datacube exposures into one, the spatial and spectral structures of the variance are reasonably flat, since the various oscillations cancel out in the combination. However, because we ignore the additional terms of the covariance matrix, the pipeline-propagated noise estimation is still wrong in terms of its absolute value.  Ideally, we should then work only with \emph{pixtable} to avoid this effect. However, this is difficult in practice because most of signal processing and visualization routines (e.g., Fast Fourier Transform) require a regularly sampled array.

To face this complex problem\footnote{Note that this variance behavior  is not specific to these observations but is currently present in all MUSE datacubes provided by the pipeline.} we have adopted a scheme to obtain a more realistic variance estimate for faint objects where the dominant source of noise is the sky. In this case the variance is a function of wavelength only. For faint objects, we will always sum up the flux over a number of spatial and spectral bins, such as (for example) a 1\arcsec\ diameter aperture to account for atmospheric seeing that extends a few \AA\ along the spectral axis. As can be seen in Fig.~\ref{fig:covariance}, the correlation impact is strongly driven by a pixel's immediate neighbors but decreases very rapidly at larger distances. The same behavior is found along the spectral axis.  Thus, if the 3D aperture size is large enough with respect to the correlation size, the variance of the aperture-summed signal should be equal to the original variance prior to resampling.

\begin{figure}[htbp]
\begin{center}
\includegraphics[width=\columnwidth]{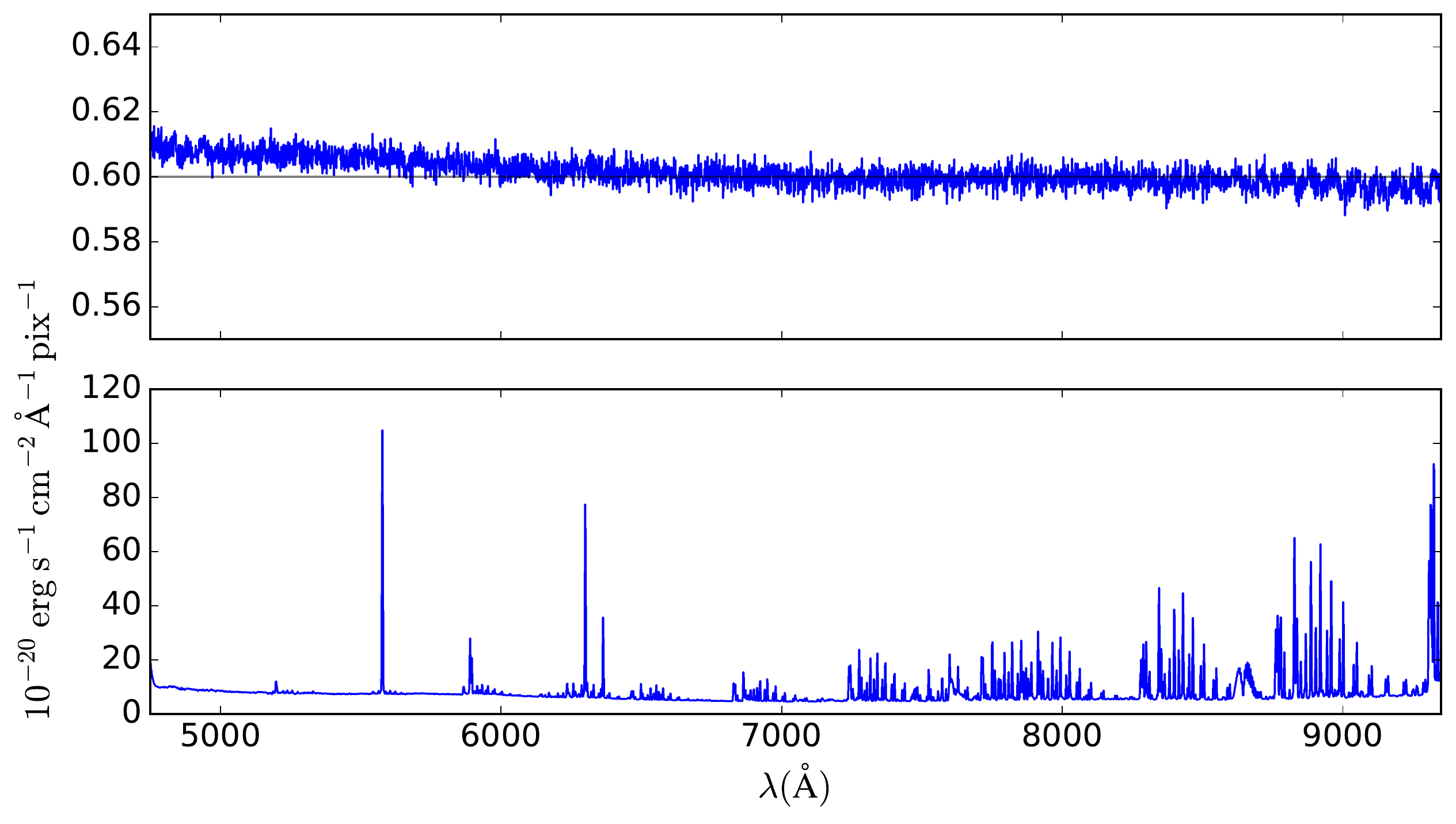}
\caption{Example of estimated standard deviation corrected for correlation effects (see text) in one exposure. 
Top: pixel-to-pixel standard deviation of the experimental noisy datacube and adopted correction factor.
Bottom: pixel-to-pixel standard deviation of a real one-exposure datacube after correcting for correlation effects.}
\label{fig:variance}
\end{center}
\end{figure}

As a test to reconstruct the original pre-resampling variances, we perform the following experiment. We start with a \emph{pixtable} that produces an individual datacube, which will later be combined with the other exposures. We fill this \emph{pixtable} with perfect Gaussian noise (with a mean of zero and a variance of 1) and then produce a datacube using the standard pipeline 3D drizzle.  As expected, the pixel-to-pixel variance of this test datacube is less than 1 because of the correlation. The actual value depends on the \emph{pixfrac} drizzle parameter related to the number of neighboring voxels which are used in the interpolation process. With our \emph{pixfrac} of 0.8, we measure a pixel-to-pixel standard deviation of 0.60 in our experimental datacube. This value is almost independant of wavelength as can be seen in Fig.~\ref{fig:variance}. The ratio $\frac{1}{0.60}$ is then the correction factor that needs to be applied to the pixel-to-pixel standard deviation.

To overcome the previously mentioned problem of noise in the pipeline-propagated variance estimator, we re-estimate the pixel-to-pixel variance directly from each datacube. We first mask the bright sources and then measure the median absolute deviation for each wavelength. The resulting standard deviation is then multiplied by the correction factor to take into account the correlations. An example is shown in Fig.~\ref{fig:variance}. Note however that this variance estimate is likely to be wrong for  bright sources which are no longer dominated by the sky noise, and thus no longer have spatially constant variances. Given the focus of the science objectives, this is not considered a major problem in this work.

\subsubsection{Exposure properties}
\begin{figure}[htbp]
\begin{center}
\includegraphics[width=\columnwidth]{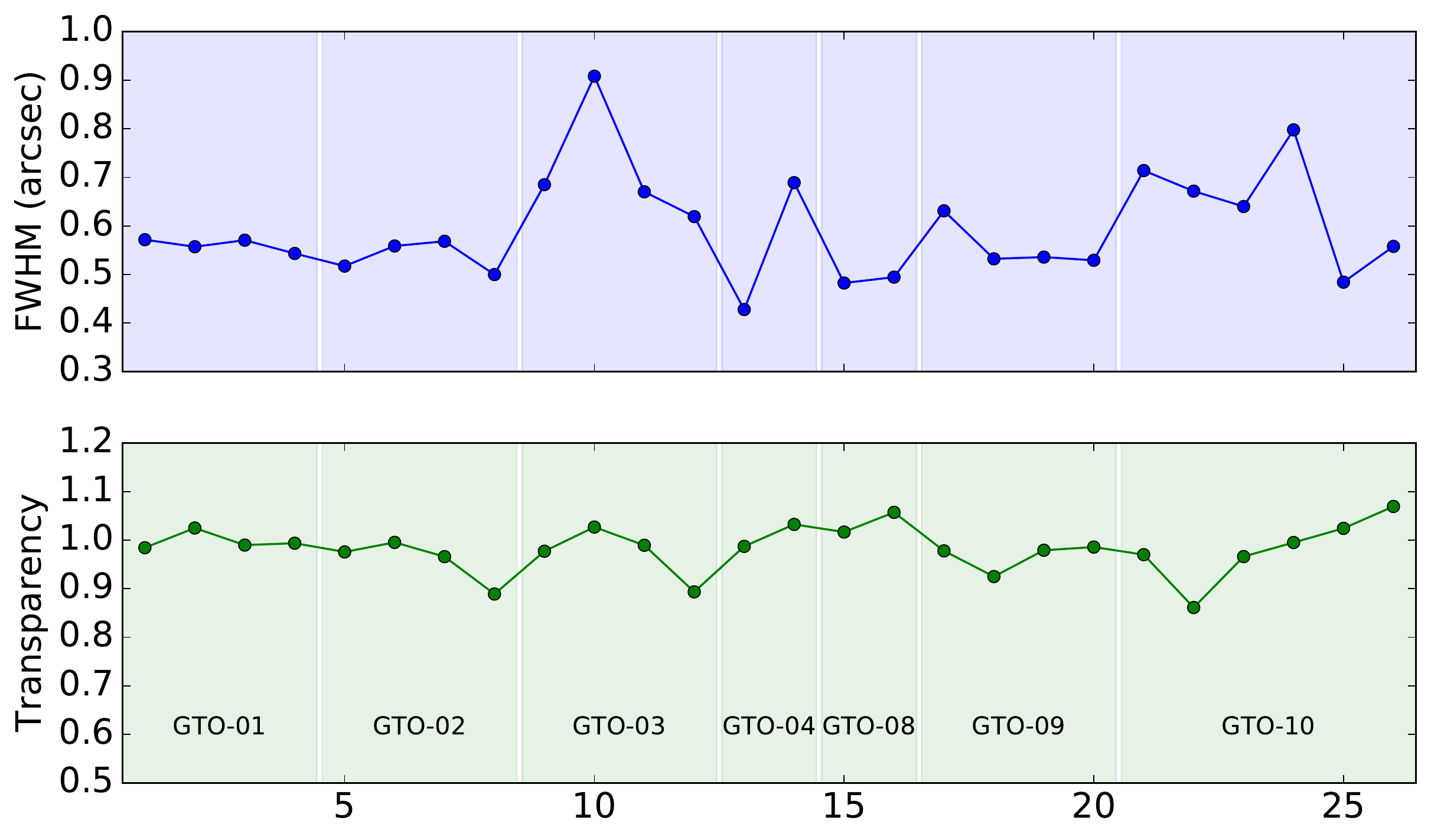}
\caption{Computed variation of FSF FWHM at 7750 \AA\ (top panel) and transparency (bottom panel) for all exposures of  the UDF-04 field obtained in seven GTO runs.}
\label{fig:exposures}
\end{center}
\end{figure}

In the final step before combining all datacubes, we evaluate some important exposure properties, such as their achieved spatial resolution and absolute photometry. We use the tool described in Sect.~\ref{sect:FSF} to derive the FWHM of the Moffat PSF fit and the photometric correction of the MUSE exposure that gives the best match with the HST broadband images. An example of the evolution of the spatial resolution and photometric properties of the UDF-04 field is given in Fig.~\ref{fig:exposures}.  The statistics of exposure properties for all fields is given in Table~\ref{tab:fields}.

Control quality pages have been produced for all 278 individual exposures displaying various images, spectra and indicators for the steps of the data reduction. They were all visually inspected, and remedy actions were performed for the identified problems. 

\begin{table*}
\caption{Observational properties of UDF fields. For each field the number of individual exposures (N) is given, along with some statistics of the FWHM (arcsec) of the estimated point spread function (FSF) at 7750\AA : the mean ($\mathrm{F_{m}}$), standard deviation ($\mathrm{F_{\sigma}}$), and min ($\mathrm{F_{min}}$) and max ($\mathrm{F_{max}}$) values. Statistics of the relative photometric properties of each field are also given: the mean ($\mathrm{P_{m}}$), standard deviation ($\mathrm{P_{\sigma}}$), and min ($\mathrm{P_{min}}$) and max ($\mathrm{P_{max}}$) values. Additionally, the fit FWHM (in arcsec) of the combined datacube FSF is given at blue ($\mathrm{F_b}$ at 4750\AA) and red ($\mathrm{F_r}$ at 9350\AA) wavelengths.}
\label{tab:fields}
\begin{center}
\begin{tabular}{r  r  rrrr rrrr  rr}
\toprule
Field & N  & $\mathrm{F_{m}}$ & $\mathrm{F_{\sigma}}$ & $\mathrm{F_{min}}$ & $\mathrm{F_{max}}$ &  $\mathrm{P_{m}}$ & $\mathrm{P_{\sigma}}$ & $\mathrm{P_{min}}$ & $\mathrm{P_{max}}$ &  $\mathrm{F_b}$  & $\mathrm{F_r}$ \\
\midrule
01 & 26 & 0.62 & 0.12 & 0.46 & 1.01 & 0.98 & 0.07 & 0.78 & 1.08 & 0.71 & 0.57 \\
02 & 28 & 0.60 & 0.11 & 0.42 & 0.82 & 1.01 & 0.02 & 0.97 & 1.06 & 0.69 & 0.56 \\
03 & 24 & 0.61 & 0.09 & 0.46 & 0.76 & 1.01 & 0.03 & 0.95 & 1.09 & 0.72 & 0.55 \\
04 & 26 & 0.59 & 0.10 & 0.43 & 0.91 & 0.98 & 0.05 & 0.86 & 1.07 & 0.72 & 0.54 \\
05 & 25 & 0.63 & 0.08 & 0.46 & 0.79 & 0.95 & 0.07 & 0.77 & 1.01 & 0.72 & 0.58 \\
06 & 24 & 0.62 & 0.06 & 0.55 & 0.78 & 0.99 & 0.03 & 0.95 & 1.07 & 0.71 & 0.56 \\
07 & 24 & 0.59 & 0.08 & 0.46 & 0.72 & 0.99 & 0.03 & 0.93 & 1.07 & 0.68 & 0.54 \\
08 & 24 & 0.63 & 0.08 & 0.43 & 0.81 & 0.98 & 0.04 & 0.86 & 1.09 & 0.72 & 0.58 \\
09 & 26 & 0.67 & 0.07 & 0.56 & 0.83 & 0.98 & 0.03 & 0.90 & 1.03 & 0.76 & 0.62 \\
10 & 51 & 0.60 & 0.08 & 0.42 & 0.77 & 1.02 & 0.04 & 0.86 & 1.09 & 0.71 & 0.55 \\
\bottomrule
\end{tabular}
\end{center}
\end{table*}

\subsection{Production of the final datacubes}

The 227 datacubes of the \mosaic\ were combined, using the estimated flux corrections computed from a comparison with the reference HST image (see section~\ref{sect:FSF}). We perform an average on all voxels, after applying a 5 sigma-clipping based on a robust median absolute deviation estimate to remove outliers. Except in the region of overlap between adjacent fields, or at the edges of the mosaic, each final voxel is created from the average of $\approx$23 voxels. The corrected variance is also propagated and an exposure map datacube is derived (see Fig.~\ref{fig:expmap}). The achieved median depth is 9.6 hours. We also save the statistics of detected outliers to check if specific regions or exposures have been abnormally rejected. The resulting datacube is saved as a 25 GB multi-extension FITS file with two extensions: the data and the estimated variance. Each extension contains $\mathrm{(n_x,n_y,n_\lambda)} = 947 \times 945 \times 3681 = 3.29 \times 10^9\, \mathrm{voxels}$.

The same process is applied to the 51 UDF-10 proper datacubes plus the 105 overlapping \mosaic\ datacubes (fields 01, 02, 04, and 05) projected onto the same grid. We note that four exposures with poor spatial resolution (FWHM > 0\farcs9)  have been removed from the combination. In this case, $\approx$74 voxels are averaged for each final voxel, leading to a median depth of 30.8 hours (Fig.~\ref{fig:expmap}). The resulting 2.9 GB datacube contains $\mathrm{(n_x,n_y,n_\lambda)} = 322  \times 323 \times 3681 = 3.8 \times 10^8\, \mathrm{voxels}$. Note that the datacubes presented in this paper have the version 0.42.

To ensure that there is no background offset, we subtract the median of each monochromatic image from each cube, after proper masking of bright sources. The subtracted offsets are small: $0.02 \pm 0.03 \times 10^{-20} \ergs$. The reconstructed white light images for the two fields, obtained simply by averaging over all wavelengths, are shown in Fig.~\ref{fig:white}.

\begin{figure}[htbp]
\begin{center}
\includegraphics[width=\columnwidth]{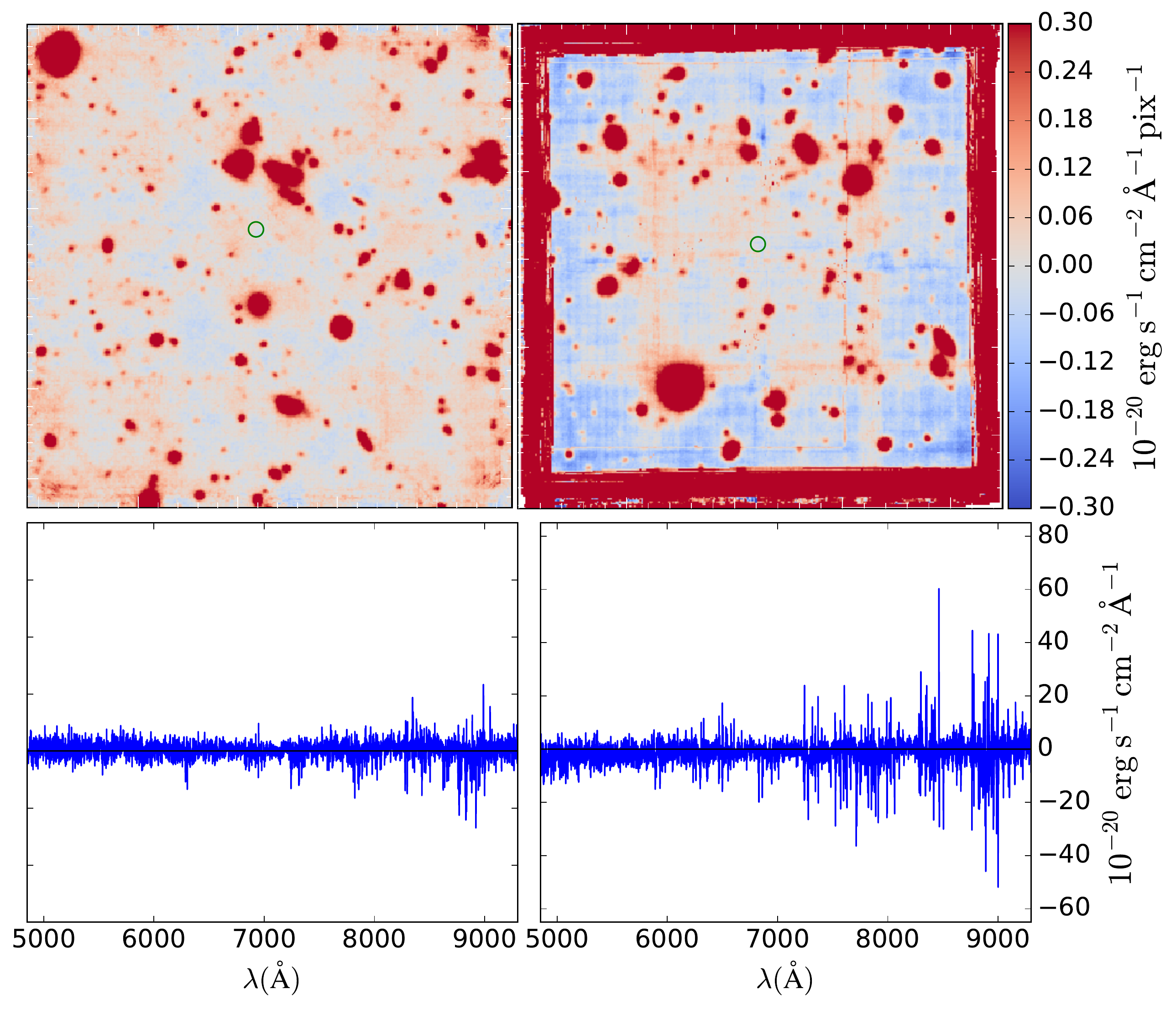}
\caption{Visual comparison between \udft\ (left) and HDFS (right) datacubes. White-light images are displayed in the top panels and examples of spectra extracted in an empty central region (green circle) are displayed in the bottom panels.}
\label{fig:comphdfs}
\end{center}
\end{figure}

\begin{figure*}[htbp]
\begin{center}
\includegraphics[width=\textwidth]{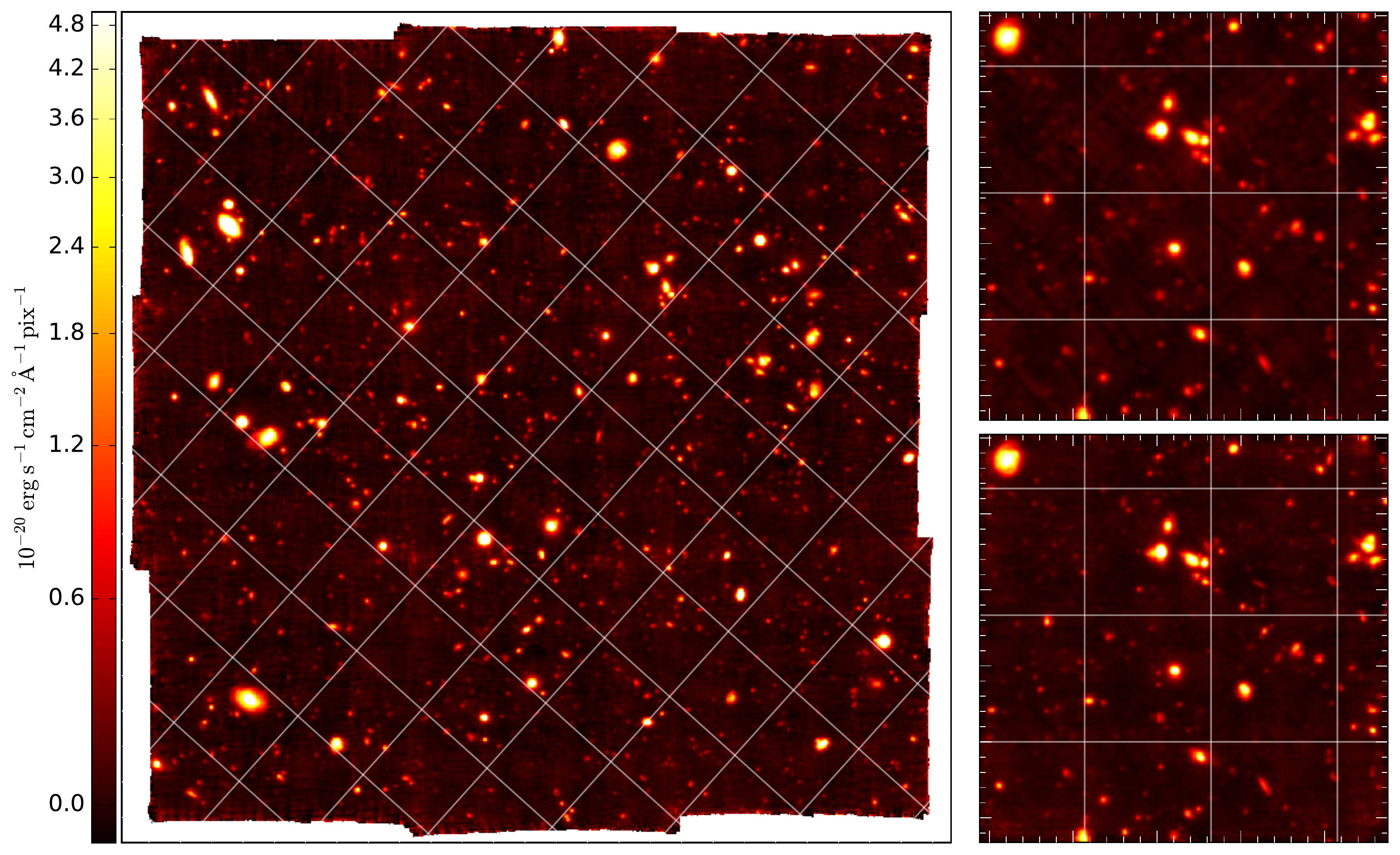}
\caption{Reconstructed white light images for the \mosaic\ (PA=-42\degr, left panel) and the \udft\ (PA=0\degr, bottom right panel). The \mosaic\ rotated and zoomed to the \udft\ field is shown for comparison in the top right panel. The grid is oriented (north up, east left) with a spacing of 20\arcsec.}
\label{fig:white}
\end{center}
\end{figure*}

To show the progress made since the HDFS publication \citep{HDFS}, we present in Fig.~\ref{fig:comphdfs} a comparison between the HDFS cube and the \udft\ cube which achieves a similar depth. There are obvious differences: the bad-edge effect present in HDFS has now disappeared, the background is much flatter in the \udft\ field, while the HDFS shows negative and positive large scale fluctuations. The sky emission line residuals are also reduced as shown in the background spectra comparison. One can also see some systematic offsets in the HDFS background at blue wavelengths which are not seen in the \udft.

\section{Astrometry and photometry}
\label{sect:astrophoto}

In the next sections we derive the broadband properties of the \mosaic\ and \udft\ datacubes by comparing their astrometry and photometry  to the HST broadband images. 

\begin{figure}[htbp]
\begin{center}
\includegraphics[width=\columnwidth]{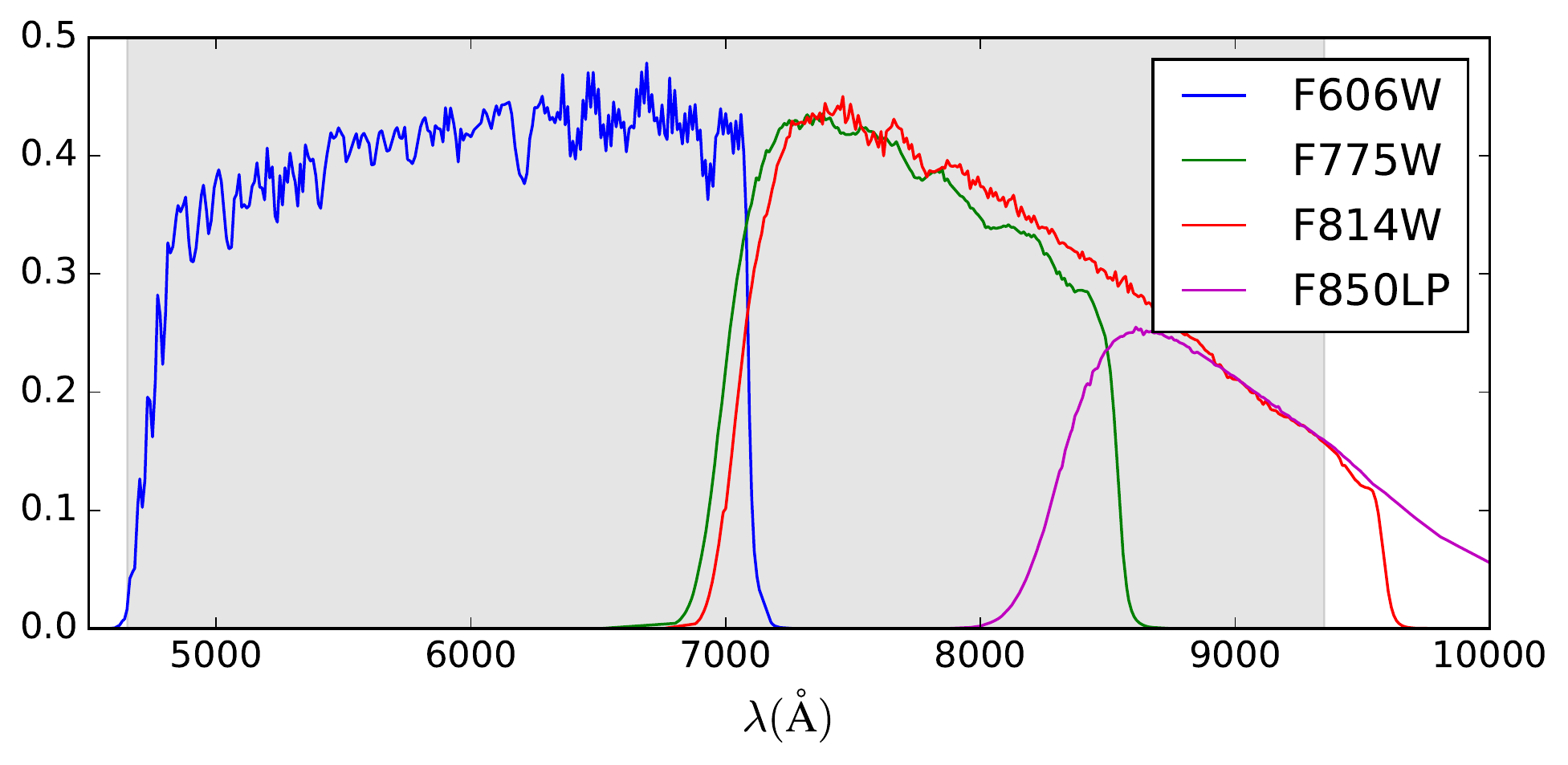}
\caption{ACS/WFC HST broadband filter response. The gray area indicates the MUSE wavelength range}
\label{fig:hstfilters}
\end{center}
\end{figure}

We derive the MUSE equivalent broadband images by a simple weighted mean of the datacubes using the ACS/WFC filter response (Fig.~\ref{fig:hstfilters}). Note that the F606W and F775W filters are fully within the MUSE wavelength range, but the two others filters (F814W and F850LP) extend slightly beyond the red limit. The corresponding HST images from the XDF data release \citep{Illingworth2013} are then broadened by convolution to match the MUSE PSF (see section~\ref{sect:FSF}) and the data are rebinned to the MUSE 0\farcs2 spatial sampling.
For the comparison with the \mosaic\ datacube, we split the HST images into the corresponding nine MUSE sub-fields in order to use the specific MUSE PSF model for each field. 

\subsection{Astrometry}
\label{Astrometry}
The \noisechisel\ software \citep{NoiseChisel} is used to build a segmentation map for each MUSE image. \noisechisel\ is a noise-based non-parametric technique for detecting nebulous objects in deep images and can be considered as an alternative to  \textsf{SExtractor} \citep{Bertin+1996}. \noisechisel\ defines "clumps" of detected pixels which are aggregated into a segmentation map. The light-weighted centroid is computed for each object and compared to the light-weighted centroid derived from the PSF-matched HST broadband image using the same segmentation map.

\begin{figure}[htbp]
\begin{center}
\includegraphics[width=\columnwidth]{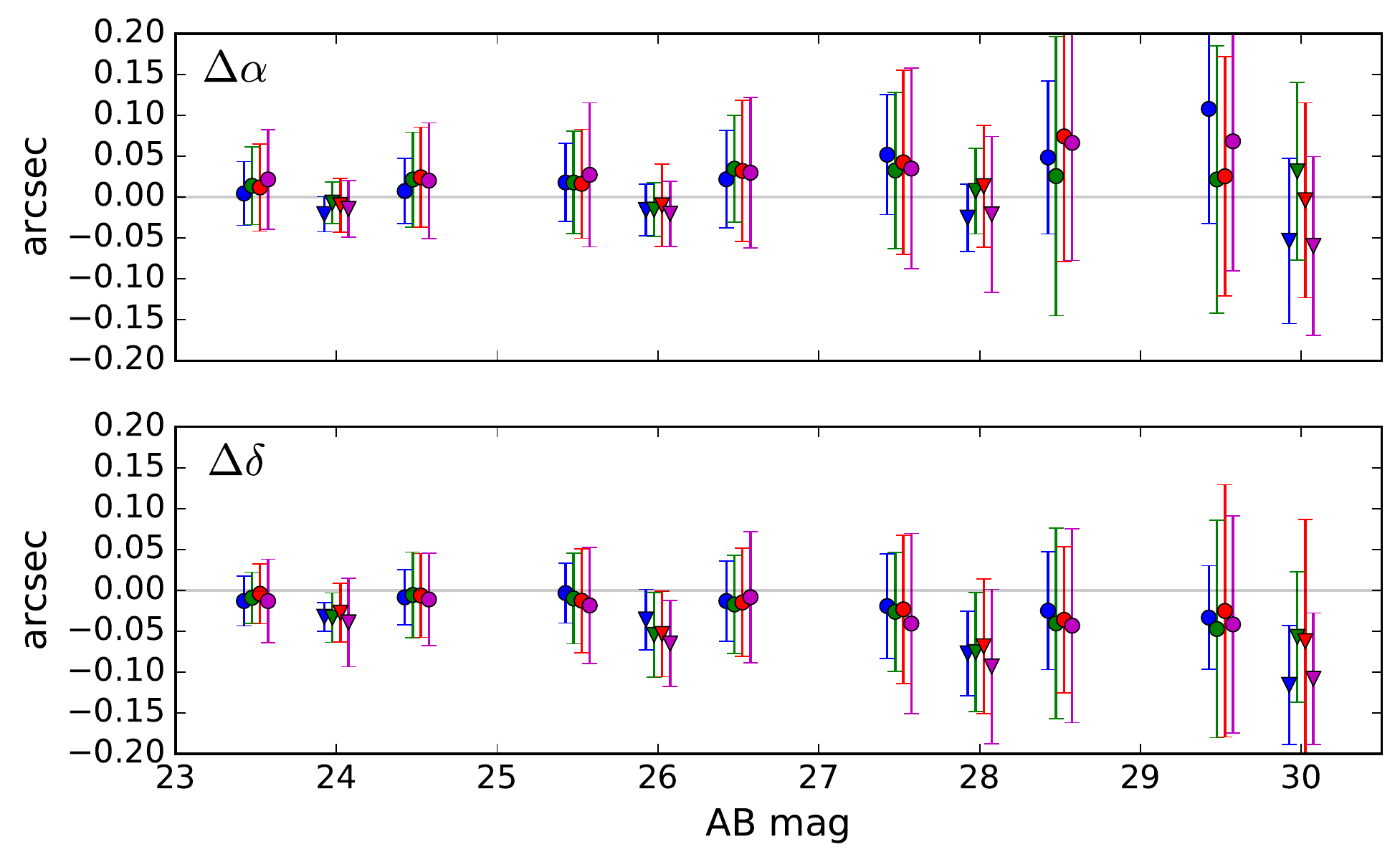}
\caption{Mean astrometric errors in $\alpha$,$\delta$ and their standard deviation in HST magnitude bins. The error bars are color coded by HST filter: blue (F606W), green (F775W), red (F814W) and magenta (F850LP). The two different symbols (circle and arrow) identify respectively the \mosaic\ and \udft\ fields. Note that \mosaic\ data are binned in 1-magnitude steps while \udft\ data points are binned over 2-magnitude steps in order to get enough points for the statistics.}
\label{fig:astrometry}
\end{center}
\end{figure}

The results of this analysis are given in Fig.~\ref{fig:astrometry} for both fields and for the four HST filters. As expected, the astrometric precision is a function of the object magnitude. There are no major differences between the filters, except for a very small increase of the standard deviation of the reddest filters.  For objects brighter than AB 27, the mean astrometric offset is less than 0\farcs035 in the \mosaic\ and less than 0\farcs030 in the \udft. The standard deviation increases with magnitude, from 0\farcs04 for bright objects up to 0\farcs15 at AB$>$29. For galaxies brighter than AB 27, we achieve an astrometric precision better than 0\farcs07 rms, i.e., 10\% of the spatial resolution.

\subsection{Photometry}
\label{Photometry}
We now compute the broadband photometric properties of our data set, using a process similar to the previous astrometric measurements. This time, however we use the \noisechisel\ segmentation maps generated from the PSF-matched HST broadband images. The higher signal-to-noise of these HST images allows us to identify more (and fainter) sources than in the MUSE equivalent image. The magnitude is then derived by a simple sum over the apertures identified in the segmentation map. We note that the background subtraction was disabled in order to measure the offset in magnitude between the two images. The process is repeated on the MUSE image using the same segmentation map and the magnitude difference saved for analysis.
Note also that we exclude the F850LP filter in this analysis because a significant fraction of its flux ($\approx$20\%) lies outside the MUSE wavelength range. 

\begin{figure}[htbp]
\begin{center}
\includegraphics[width=\columnwidth]{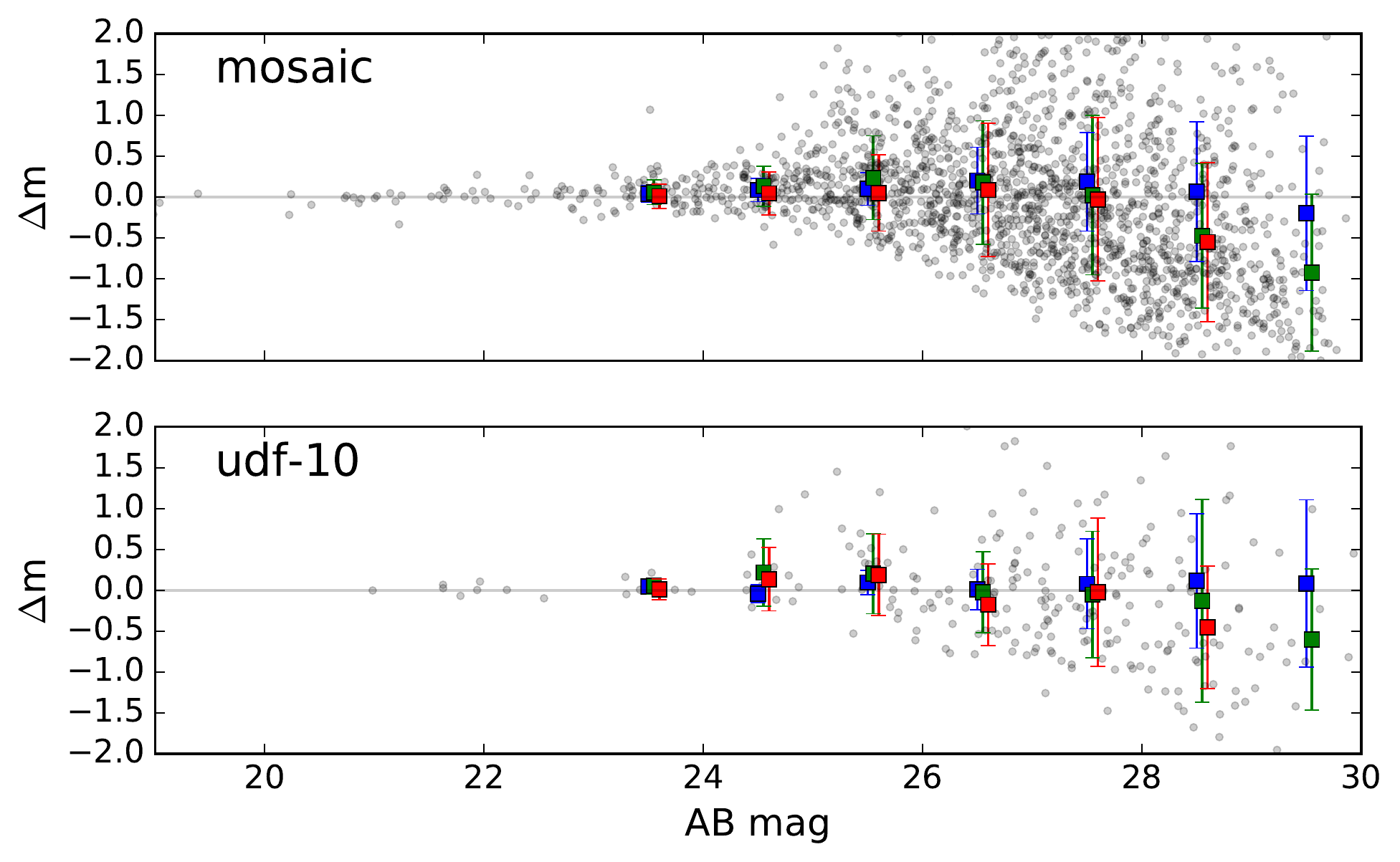}
\caption{Differences between MUSE and HST AB broadband magnitudes. The gray points show the individual measurements for the F775W filter. The mean AB photometric errors and their standard deviations in HST magnitude bins are shown as error bars, color coded by HST filter: blue (F606W), green (F775W) and red (F814W). Top and bottom panels respectively show the \mosaic\ and \udft\ fields.}
\label{fig:photometry}
\end{center}
\end{figure}

The result of this comparison is shown in Fig.~\ref{fig:photometry}. The MUSE magnitudes match their HST counterparts well, with little systematic offset up to AB 28 ($\mathrm{\Delta m< 0.2}$). For fainter objects, MUSE tends to under-estimate the flux with an offset more prominent in the red filters. The exact reason for this offset is not known but it may be due to some systematic left over by the sky subtraction process. As expected, the standard deviation increases with magnitude and is larger in the red than in the blue, most probably because of sky residuals. For example, the \mosaic\ scatter is 0.4 magnitudes in F606W at 26.5 AB, but is a factor of two larger in the F775W and F814W filters at the same magnitude. By comparison, the deeper \udft\ datacube achieves better photometric performance with a measured rms that is 20-30\% lower than in the \mosaic. 

\section{Spatial and spectral resolution}
\label{sect:psf}

A precise knowledge of the achieved spatial and spectral resolution is key  for all subsequent analysis of the data. For ground based observations where the exposures are obtained under various, and generally poorly known,  seeing conditions,  knowledge of the spatial PSF is also important for each individual exposure. For example, exposures with bad seeing will add more noise than signal for the smaller sources and should be discarded in the final combination of the exposures. Note that the assessment of the spatial PSF for each individual exposure does not need to be as precise as for the final combined datacube.

The spectral resolution is not impacted by the change of atmospheric conditions and the instrument is stable enough to avoid the need of a spectral PSF evaluation for each individual exposure. However, good knowledge of the spectral resolution in the final datacube is also required.

In the next sections we describe the results and the methods used to derive these PSFs. To distinguish between the spectral and spatial axes, we name the spectral line spread function and the field spatial point spread function, LSF and FSF, respectively.

\subsection{Spatial Point Spread Function (FSF)}
\label{sect:FSF}
In the ideal case of a  uniform FSF over the field of view, its evaluation is straightforward if one has a bright point source in the field. If we assume a Gaussian shape, then only one parameter, the FWHM, fully characterizes the FSF.
In our case we are not far from this ideal case, because the MUSE field  is quite small with respect to the telescope field of view and its image quality ($\sim$0\farcs2) is much better than the seeing size. However, given the long wavelength range of MUSE, one cannot neglect the wavelength dependence of the seeing. For the VLT's large aperture, a good representation of the atmospheric turbulence is given by \cite{Tokovinin2002} in the form of a finite outer scale von Karman turbulence model. It predicts a nearly linear decrease of FWHM with respect to wavelength, with the slope being a function of the atmospheric seeing and the outer scale turbulence. 

During commissioning, a detailed analysis of the MUSE FSF showed that it was
very well modeled by a Moffat circular function $[1 - (r/\alpha)^\beta]^{-\frac{1}{2}}$
with $\beta$ constant and a linear variation of $\alpha$ with wavelength. The
same parametrisation was successfully used in the HDFS study \citep{HDFS}
using the brightest star (R=19.6) in the field. However, most of MUSE UDF
fields do not have such a bright star and the majority of our fields have no
star with $R<23$ at all.

Fortunately, broadband HST images of the UDF exist for many wavelengths. In
particular, as shown in Fig.~\ref{fig:hstfilters}, the wavelength coverage of four HST imaging filters, F606W, F775W,
F814W and F850LP falls entirely or partially within the MUSE
wavelength range (4750-9350 \AA). If one of these images is convolved with the
MUSE FSF, and the equivalent MUSE image is convolved with the HST FSF, then
the resulting images should end up with the same combined FSF. Thus, the
similarity of HST and MUSE images that have been convolved with models of each
other's FSFs, can be used to determine how well those models match the data.

\newcommand{\src}{\mathbf{s}}
\newcommand{\imagem}{\mathbf{d_m}}
\newcommand{\imageh}{\mathbf{d_h}}
\newcommand{\fsfm}{\mathbf{\psi_m}}
\newcommand{\fsfh}{\mathbf{\psi_h}}
\newcommand{\noisem}{\mathbf{n_m}}
\newcommand{\noiseh}{\mathbf{n_h}}
\newcommand{\modelm}{\fsfm^{\prime}}
\newcommand{\modelh}{\fsfh^{\prime}}

In the following equations, suffixes of $m$ and $h$ are used to distinguish
between symbols associated with the MUSE and HST images, respectively.
Equation~\ref{eqn:museimage} models a MUSE image ($\imagem$) as a perfect
image of field sources ($\src$) convolved with the MUSE FSF ($\fsfm$), summed
with an image of random noise ($\noisem$). Equation~\ref{eqn:hstimage} is the
equivalent equation for an HST image of the same region of the sky, but this
time convolved with the HST FSF ($\fsfh$), and summed with a different
instrumental noise image, $\noiseh$.

\begin{eqnarray}
\label{eqn:museimage}
 \imagem &=& \src \ast \fsfm + \noisem, \\
\label{eqn:hstimage}
 \imageh &=& \src \ast \fsfh + \noiseh.
\end{eqnarray}

\noindent When these images are convolved with estimated models of each
other's FSF, the result is as follows:

\begin{eqnarray}
 \imagem \ast \modelh &=& \src \ast \fsfm \ast \modelh + \noisem \ast \modelh,\\
 \imageh \ast \modelm &=& \src \ast \fsfh \ast \modelm + \noiseh \ast \modelm.
\end{eqnarray}

\noindent In these equations, $\modelm$ and $\modelh$ denote models
of the true MUSE and HST FSF profiles, $\fsfm$ and $\fsfh$. The following
equation shows the difference between these two equations.

\begin{equation}
\label{eqn:imagediff}
 \imagem \ast \modelh - \imageh \ast \modelm = \src \ast (\fsfm \ast \modelh -
 \fsfh \ast \modelm) + (\noisem \ast \modelh - \noiseh \ast \modelm)
\end{equation}

\noindent The magnitude of the first bracketed term can be minimized by
finding accurate models of the MUSE and HST FSFs. However, this is not a unique
solution, because the magnitude can also be minimized by choosing accurate
models of the FSF profiles that have both been convolved by an arbitrary
function. To unambiguously evaluate the accuracy of a given model of the MUSE
FSF, it is thus necessary to first obtain a reliable independent estimate of
the HST FSF.  This can be achieved by fitting an FSF profile to bright stars
within the wider HST UDF image.

Minimizing the first of the bracketed terms of equation~\ref{eqn:imagediff}
does not necessarily minimize the overall equation. The noise contribution
from the second of the bracketed terms decreases steadily with increasing FSF
width, because of the averaging effect of wider FSFs, so the best-fit MUSE FSF
is generally slightly wider than the true MUSE FSF.  However provided that the
image contains sources that are brighter than the noise, the response of the
first bracketed term to an FSF mismatch is greater than the decrease in the
second term, so this bias is minimal.

In summary, with a reliable independent estimate of the HST FSF\footnote{In practice we compute the Moffat fit for a few bright stars in the field for each HST filter.}, a good
estimate of the MUSE FSF can be obtained by minimizing the magnitude of
Eq.~\ref{eqn:imagediff}, as a function of the model parameters of the
FSF. In practice, to apply this equation to digitized images, the pixels of
the MUSE and HST images must sample the same positions on the sky, have the
same flux calibration, and have the same spectral response. A MUSE image of
the same spectral response as an HST image can be obtained by performing a
weighted mean of the 2D spectral planes of a MUSE cube, after weighting each
spectral plane by the integral of the HST filter curve over the bandpass of
that plane.

HST images have higher spatial resolutions than MUSE images, so the HST image
must be translated, rotated and down-sampled onto the coordinates of the MUSE
pixel grid. Before down-sampling, a decimation filter must be applied to the
HST image, both to avoid introducing aliasing artifacts, and to remove noise
at high spatial frequencies, which would otherwise be folded to lower spatial
frequencies and reduce the signal-to-noise ratio (S/N) of the downsampled image.  The model of the HST
FSF must then be modified to account for the widening effect of the
combination of the decimation filter and the spatial frequency response of the
widened pixels.

Once the HST image has been resampled onto the same pixel grid as the MUSE
image, there are usually still some differences between the relative positions
of features in the two images, due to derotator wobble and/or telescope pointing errors. Similarly,
after the HST pixel values have been given the same flux units as the MUSE
image, the absolute flux calibration factors and offsets of the two images are not
precisely the same. To correct these residual errors, the MUSE FSF fitting
process has to simultaneously fit for position corrections and calibration
corrections, while also fitting for the parameters of the MUSE FSF.

The current fitting procedure does not attempt to correct for rotational
errors in the telescope pointing, or account for focal plane
distortions. Focal plane distortions appear to be minimal for the HST and MUSE
images, and only two MUSE images were found to be slightly rotated relative to
the HST images. In the two discrepant cases, the rotation was measured by
hand, and corrected before the final fits were performed.

As described earlier, the FSF of a MUSE image is best modeled as a Moffat
function.  Moffat functions fall off relatively slowly away from their central
cores, so a large convolution kernel is needed to accurately convolve an image
with a MUSE FSF. Convolution in the image plane is very slow for large
kernels, so it is more efficient to perform FSF convolutions in the Fourier
domain. Similarly, correcting the pointing of an image by a fractional number
of pixels in the image domain requires interpolation between pixels, which is
slow and changes the FSF that is being measured.  In the Fourier domain, the
same pointing corrections can be applied quickly without interpolation, using
the Fourier-transform shift theorem. For these reasons, the FSF fitting
process is better performed entirely within the Fourier domain, as described
below.

Let $b$ and $\gamma$ be the offset and scale factor needed to match the HST
image photometry to that of the MUSE image, and let $\epsilon$ represent the
vector pointing-offset between the HST image and the MUSE image. When the left
side of equation~\ref{eqn:imagediff} is augmented to include these
corrections, the result is the left side of the following equation:

\newcommand{\ftarrow}{\stackrel{\mbox{\scriptsize\it FT}}{\rightarrow}}
\newcommand{\Imagem}{\mathbf{D_m}}
\newcommand{\Imageh}{\mathbf{D_h}}
\newcommand{\Fsfm}{\mathbf{\Psi_m}}
\newcommand{\Fsfh}{\mathbf{\Psi_h}}
\newcommand{\Modelm}{\Fsfm^{\prime}}
\newcommand{\Modelh}{\Fsfh^{\prime}}

\begin{equation}
\label{eqn:ftdiff}
 \imagem \ast \modelh - \gamma\imageh \ast \modelm \ast \Delta(\mathbf{p}-\epsilon) + b
 \ftarrow  \Imagem \Modelh - \gamma\Imageh \Modelm e^{-i 2\pi \mathbf{f}\epsilon} + b
\end{equation}

\noindent Note that the pointing correction vector ($\epsilon$) is applied by
convolving the HST image by the shifted Dirac delta function,
$\Delta(\mathbf{p}-\epsilon)$, where $\mathbf{p}$ represents the array of
pixel positions.

The right side of equation~\ref{eqn:ftdiff} is the Fourier transform of
the left side, with $\imageh \ftarrow \Imageh$, $\imagem \ftarrow \Imagem$,
$\fsfm \ftarrow \Fsfm$ and $\fsfh \ftarrow \Fsfh$. The spatial frequency
coordinates of the Fourier transform pixels are denoted
$\mathbf{f}$. Note that all of the convolutions on the left side of the
equation become simple multiplications in the Fourier domain. The exponential
term results from the Fourier transform shift theorem, which, as shown above, is
equivalent to an image-plane convolution with a shifted delta function.

The fitting procedure uses the Levenberg-Marquardt non-linear least-squares
method to minimize the sum of the squares of the right side of
equation~\ref{eqn:ftdiff}. The procedure starts by obtaining the discrete
Fourier transforms, $\Imagem$, $\Imageh$, and $\Modelh$ using the Fast Fourier
Transform (FFT) algorithm. Then for each iteration of the fit, new trial
values are chosen for $\gamma$, $b$, $\epsilon$ and the model parameters of
the MUSE FSF, $\fsfm$. There is no analytic form for the Fourier transform of
a 2D Moffat function, so at each iteration of the fit, the trial MUSE FSF must
be sampled in the image plane, then transformed to the Fourier domain using an
FFT. It is important to note that to avoid significant circular convolution,
all images that are passed to the FFT algorithm should be zero padded to add
margins that are at least as wide as the core of the trial Moffat profiles and
the maximum expected pointing correction.

MUSE and HST images commonly contain pixels that have been masked due
to instrumental problems, or incomplete field coverage. In addition,
areas of the images that contain nearby bright stars should be masked
before the FSF procedure, because the effect of the proper motion of
these stars is often sufficiently large between the epochs of MUSE and
HST observations, to make it impossible to line up the stars without
misaligning other sources.  Since the FFT algorithm cannot cope with
missing samples, masked pixels must be replaced by a finite value.
Here, we choose a replacement value of zero, since this choice makes
the fit of the calibration scale factor ($\gamma$) insensitive to the
existence of missing pixels. However, a contiguous region of
zero-valued pixels can fool the algorithm, making it think the region
(which is significantly different from its surroundings) is a real
feature to be fit. To avoid this, we first subtract the median flux
value from each image before replacing the masked pixels with zero.
This decreases the contrast around the masked pixels, increasing the
probability that they will blend into the background and be ignored by
the fitting routine.  The median-subracted flux value is saved and
folded into the fit of the background offset parameter ($b$).

\begin{figure}[htbp]
\begin{center}
\includegraphics[width=\columnwidth]{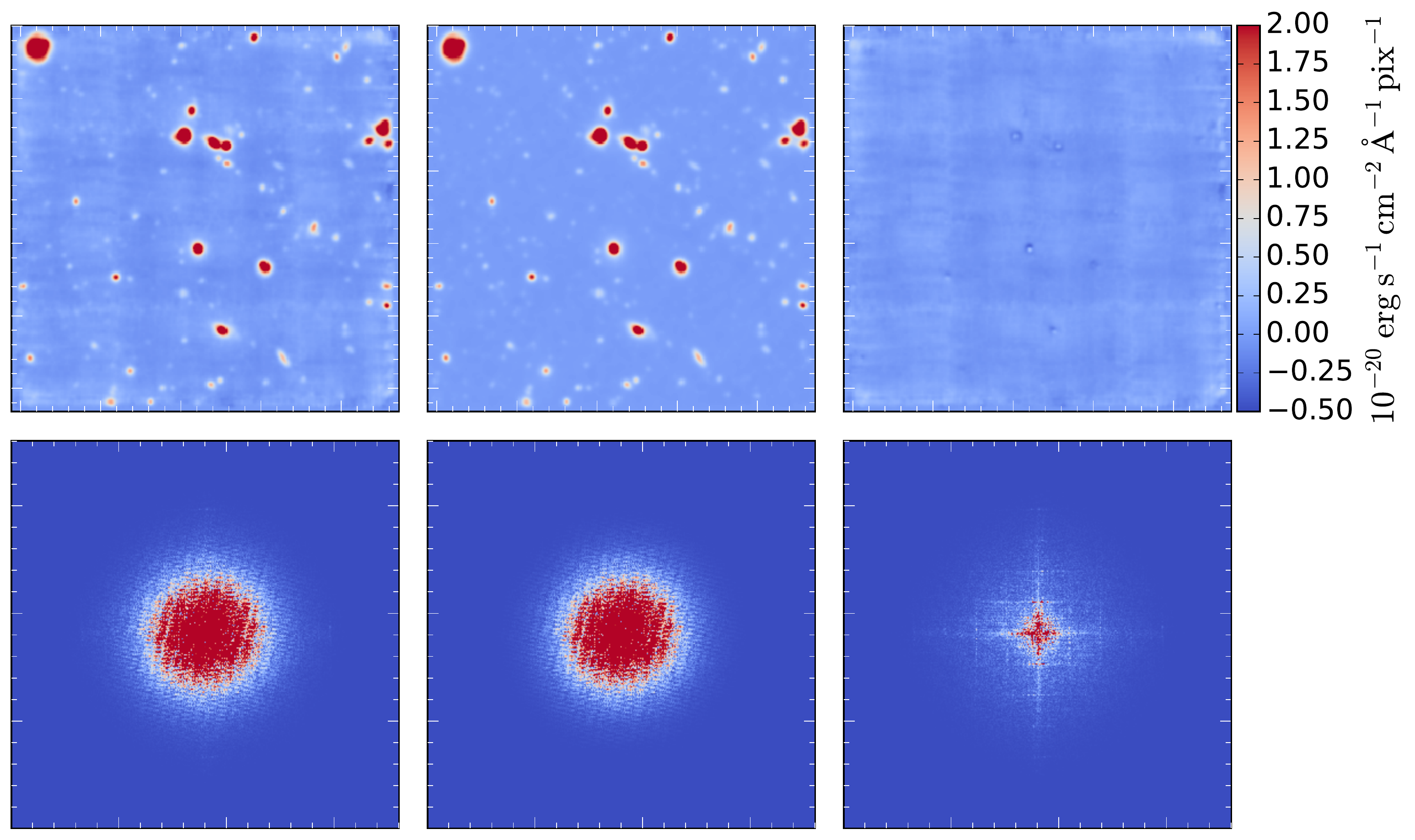}
\caption{An example demonstrating the success of the FSF fitting
  technique. The upper left panel shows the \udft\ data, rescaled by
  the equivalent HST F775W broadband filter.  The upper middle panel
  shows the corresponding HST F775W image, after it has been resampled
  onto the pixel grid of the MUSE image and convolved with the
  best-fit MUSE FSF. The upper right panel presents the residual of
  these two images, showing that only the instrumental background of
  the MUSE image remains. The lower panels show the corresponding
  images in the Fourier space where the fit is performed.}

\label{fig:fitimage}
\end{center}
\end{figure}

\begin{figure}[htbp]
\begin{center}
\includegraphics[width=\columnwidth]{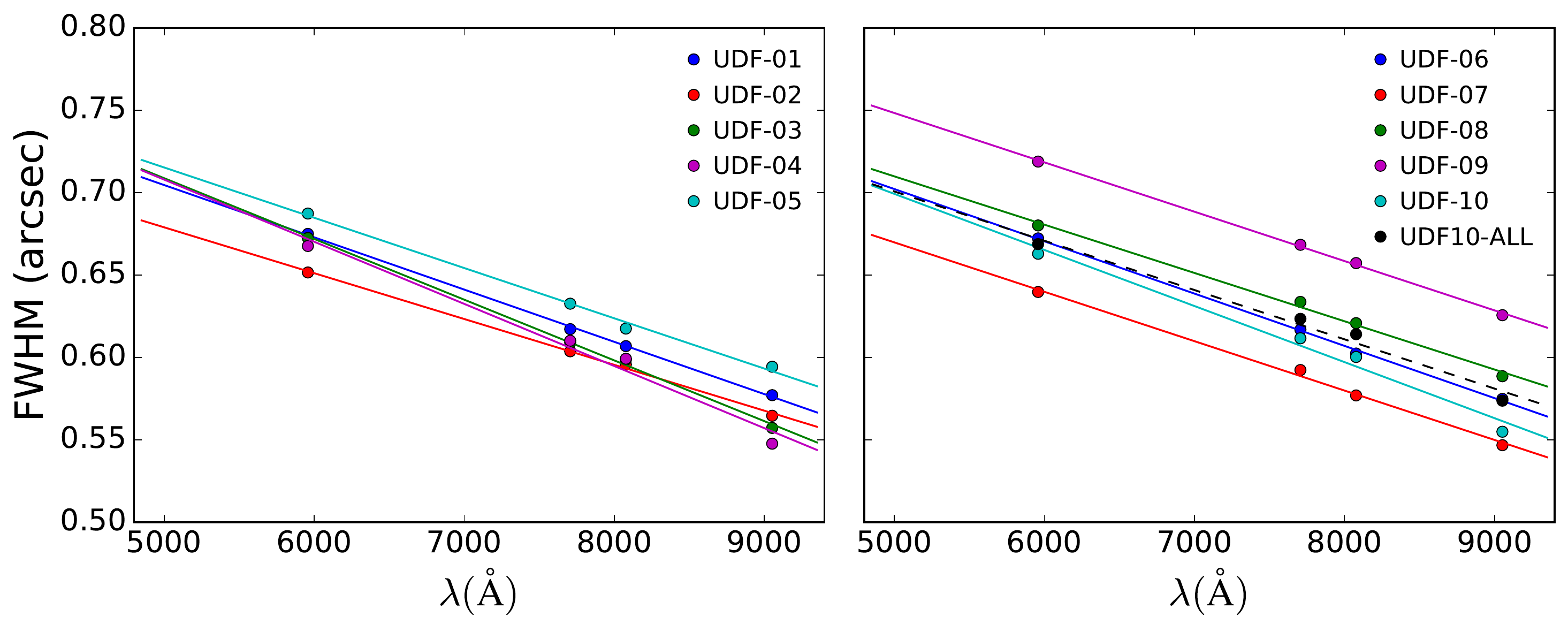}
\caption{FSF Fitting results for all \mosaic\ and \udft\ fields. For each field, 4 fit MOFFAT FWHMs corresponding to 4 HST filters (F606W, F775W, F814W, F850LP) are displayed, together with the linear fit. The UDF10-ALL is for the combined depth of the \udft\ field and its associated mosaic fields (1, 2 ,4 and 5).}
\label{fig:fsffields}
\end{center}
\end{figure}

Figure~\ref{fig:fitimage} shows an example of how well this method works in
practice and Figure~\ref{fig:fsffields} displays the fitting results obtained for all fields.  The fit values for the combined datacubes of each field are given in Table~\ref{tab:fields}.

\subsection{Spectral Line Spread Function (LSF)}
\label{sect:LSF}
To measure the LSF, we produce combined datacubes similar to the
\udft\ and \mosaic\ datacubes but without including the sky
subtraction. From these, we calculate the LSF using 19 groups of 1-10
sky lines. While the lines within each group are unresolved at the
MUSE spectral resolution, they must be accounted for to construct a
proper LSF model. For each group we used the CAMEL software (see
\citealt{Epinat2012, Contini2016} for a description of the software)
to fit a Gaussian to each line, keeping the relative position and FWHM
identical for all lines in the group. This is performed over all
spaxels  in the datacube, after applying a Gaussian spatial smoothing kernel of
  0\farcs4 FWHM to improve the S/N of the faint sky lines.

\begin{figure}[htbp]
\begin{center}
\includegraphics[width=\columnwidth]{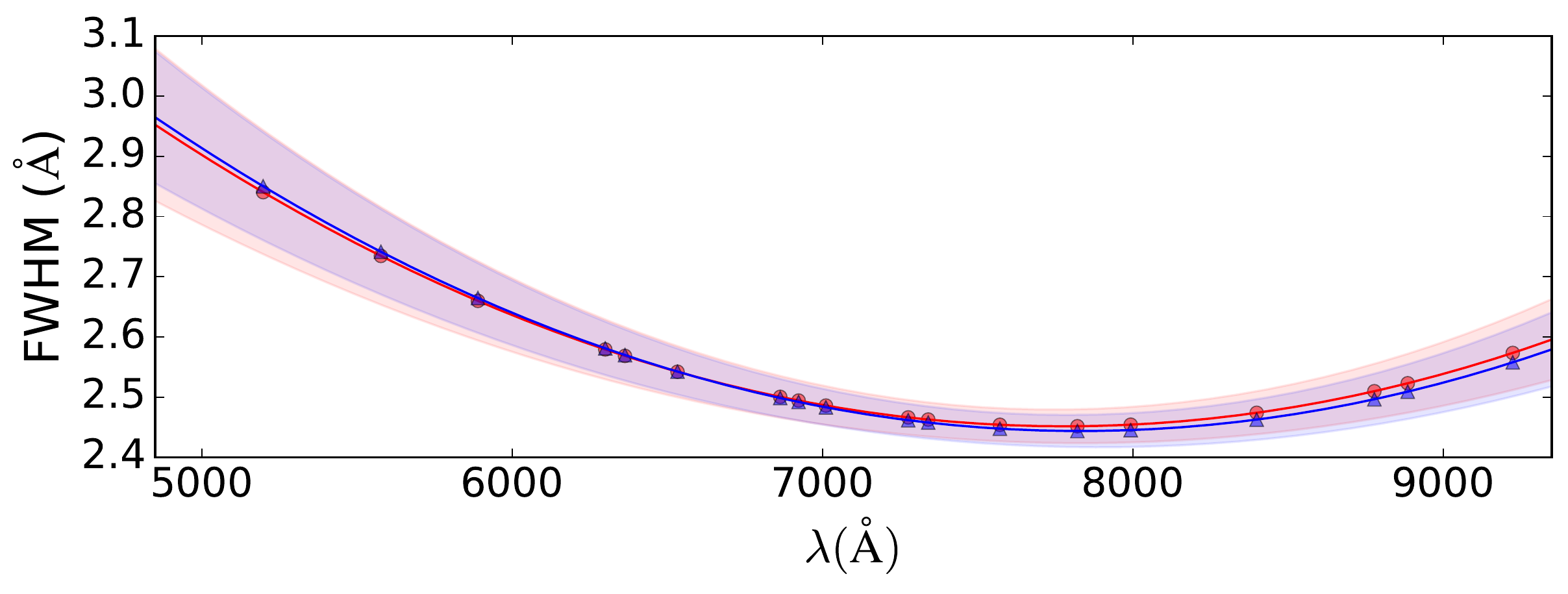}
\caption{Measured mean LSF FWHM on the \udft\ (blue line) and \mosaic\ (red line) datacubes. The symbols represent measured values while the solid line represents the polynomial fit. The shaded area shows the $\mathrm{\pm 1 \sigma}$ spatial standard deviation.}
\label{fig:lsf}
\end{center}
\end{figure}

We show the mean and standard deviation of the resulting FWHM as a
function of wavelength in Fig.~\ref{fig:lsf}. Note that there is, as
expected, little difference between the \udft\ and
\mosaic\ datacubes. The FWHM of the modeled LSF varies smoothly with
wavelength, ranging from 3.0 \AA\ (at the blue end) to 2.4 \AA\ (at
7500 \AA). It remains largely constant over the field of view, with an
average standard deviation of 0.05 \AA.  The FWHM variations as a function
of wavelength $\mathrm{F(\lambda)}$ (in \AA) are best described by polynomial
functions:
\begin{equation}
\mathrm{F_{mosaic}}(\lambda) =  5.835\;  10^{-8} \lambda^2 -9.080\;  10^{-4} \lambda + 5.983
\end{equation}

\begin{equation}
\mathrm{F_{udf10}}(\lambda) = 5.866\;  10^{-8} \lambda^2 -9.187\; 10^{-4} \lambda + 6.040
\end{equation}

We note that the true LSF shape is not actually Gaussian, but instead
more square in shape. The simple Gaussian model is However a good
approximation for most usage.

\section{Noise properties and limiting flux}
\label{sect:noise}

\subsection{Noise properties}
\begin{figure}[htbp]
\begin{center}
\includegraphics[width=\columnwidth]{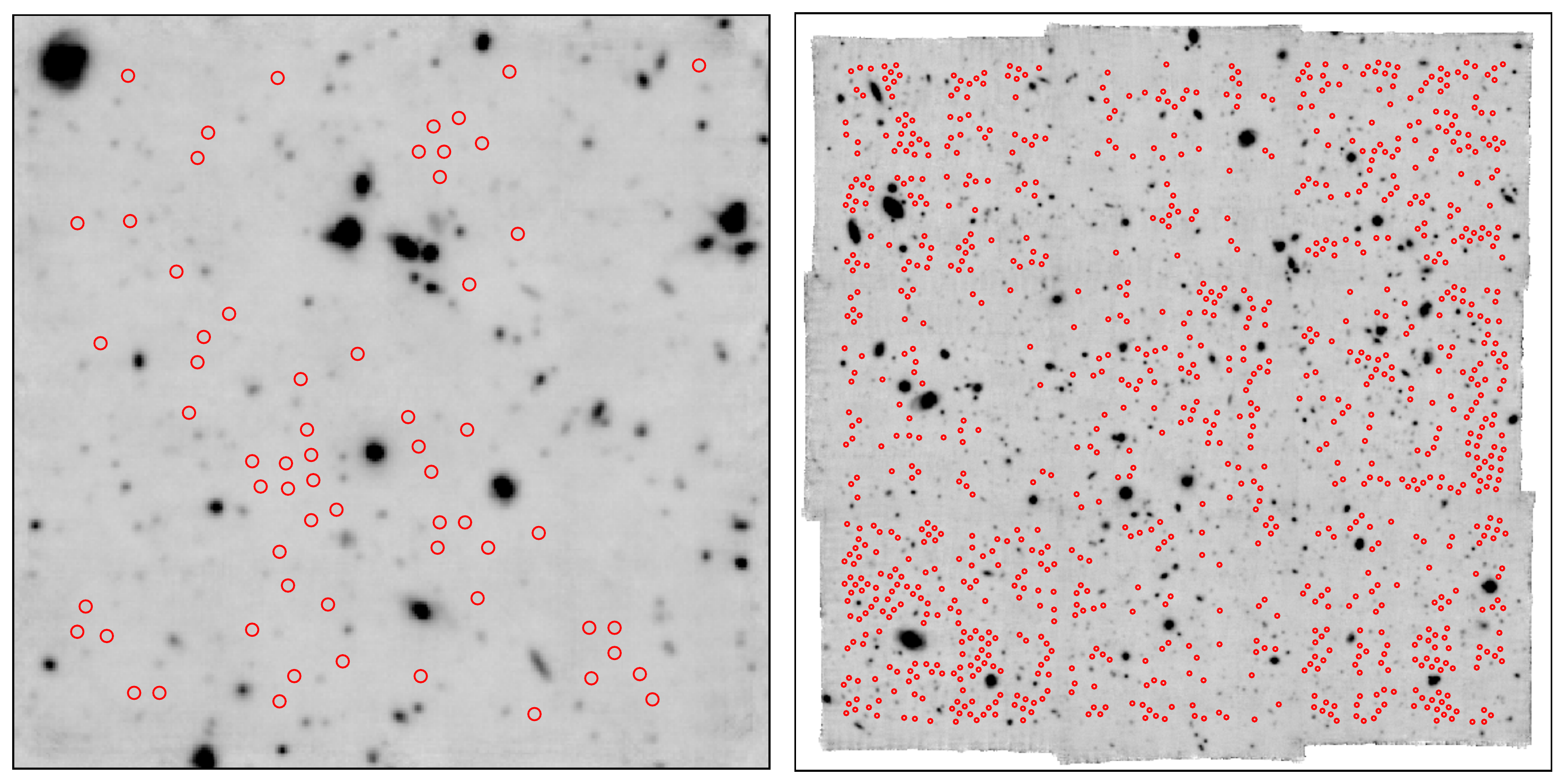}
\caption{Selected apertures used to evaluate the variance in empty regions of the \udft\ (left panel) and \mosaic\ (right panel) datacubes.}
\label{fig:apertures}
\end{center}
\end{figure}

The empirical procedure described in Sect.~\ref{sect:variance} should correct the variance estimate for the correlation added by the 3D drizzle interpolation process. We thus expect the propagated variance of the final datacubes to be correct in that respect. To check that this is indeed the case, we estimate the variance from a set of empty regions in the datacubes, selected to have similar integration time using the exposure maps shown in Fig.\ref{fig:expmap}. For the \udft\ field, we select 63 circular apertures of 1\arcsec\ diameter in regions with $31\pm0.3$ hours of integration time. In the \mosaic\ we select 991 similar apertures in regions with  $9.9 \pm 0.4$ hours of integration time. The locations of all selected regions are shown in Fig~\ref{fig:apertures}.

\begin{figure}[htbp]
\begin{center}
\includegraphics[width=\columnwidth]{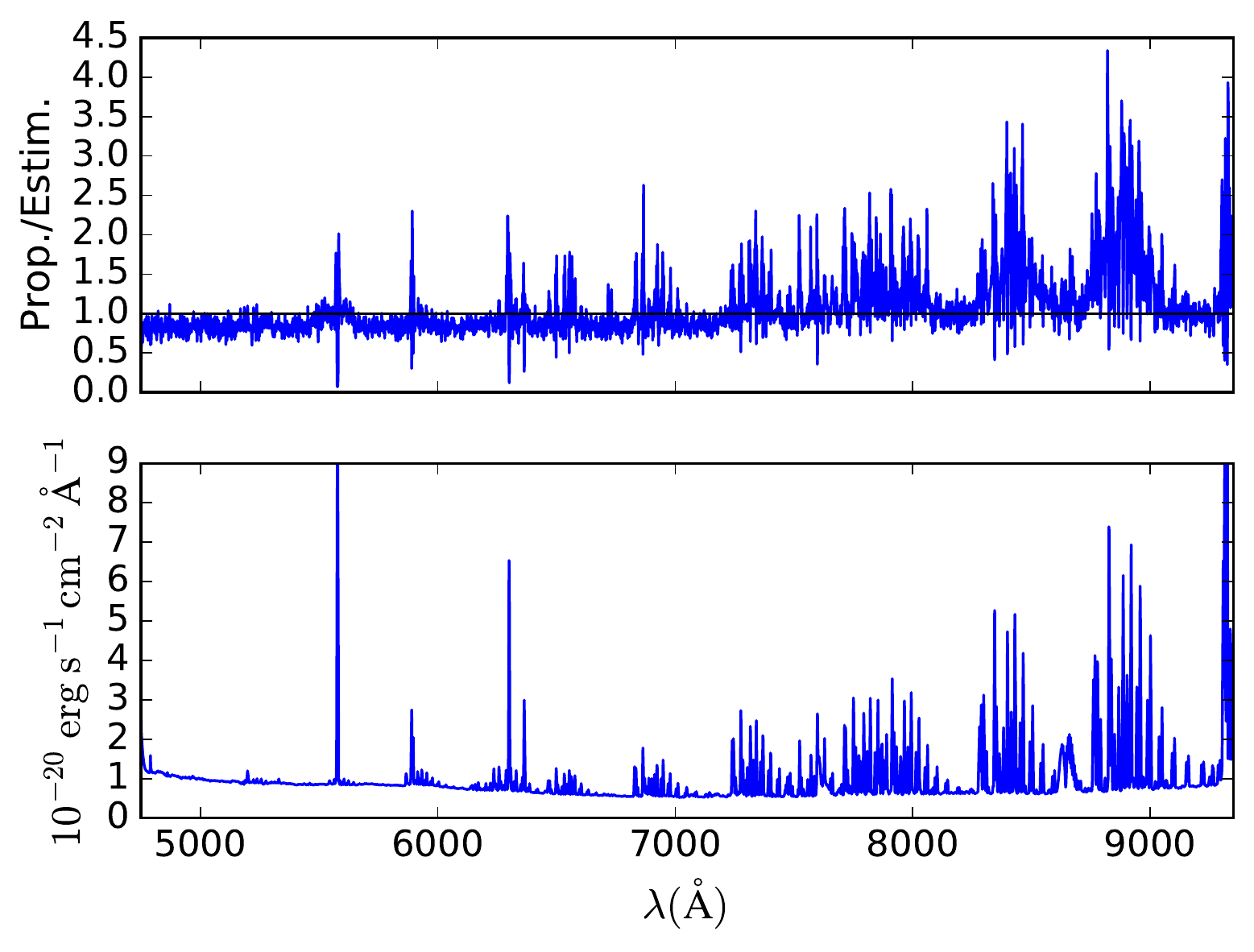}
\caption{Lower panel: Median value of the propagated noise standard deviation  for the 63 selected 1\arcsec\ diameter apertures (see text). Top panel: Ratio of the propagated to the estimated standard deviations.}
\label{fig:noise_ratio}
\end{center}
\end{figure}

We calculate the corresponding propagated variance spectrum by taking the median of the stack of all apertures.  The spectrum generated from the \udft\ field, along with the ratio between this standard deviation and the estimated standard deviation calculated in  Sect.~\ref{sect:variance} are shown in Fig.~\ref{fig:noise_ratio}.  As expected, the computed ratio is around unity\footnote{According to Fig.~\ref{fig:noise_ratio} the propagated standard deviation underestimate the compute values by $\approx$10-15\% but we did not attempt to correct for this small offset.} and constant with wavelength, showing that the propagated variance is now a good representation of the true variance within an aperture. In the top panel of Fig.~\ref{fig:noise_ratio} it is clear that there is a mismatch between the the estimated and propagated standard deviation at wavelengths that contain bright sky emission lines. The difference is due to the PCA ZAP process and discussed in detail in section 5 of \cite{ZAP}: when ZAP is applied to the individual datacubes (see Sect.~\ref{sect:skysub}) it tends to preferentially remove the strongest systematic signals left by the imperfect sky subtraction at the locations of the bright sky lines. For the brightest OH lines this results in an over-fitting of the noise which then biases the estimated variance. In that respect, the propagated variance is a better representation of the true variance. The same behavior is found for the \mosaic\ datacube.

\begin{figure}[htbp]
\begin{center}
\includegraphics[width=\columnwidth]{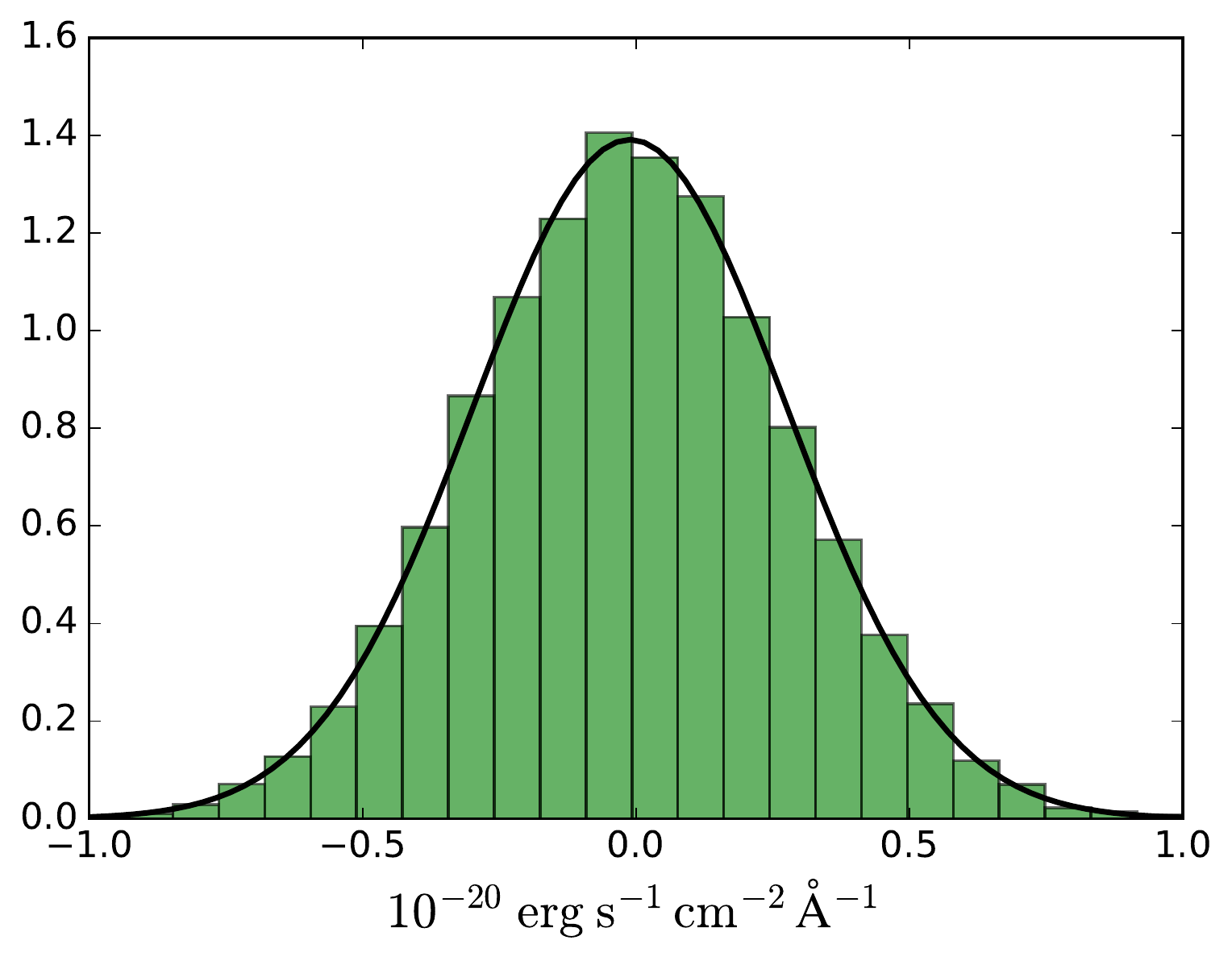}
\caption{Example of a normalized data histogram derived from an empty aperture of 1\arcsec\ diameter at 7125\AA\ in the \udft\ datacube. The solid line displays the best fit Normal PDF with a standard deviation of  $0.33 \times 10^{-20}\ergs$.}
\label{fig:histogram}
\end{center}
\end{figure}

Using the set of empty apertures we are also able to investigate the noise probability density distribution. A normal test \citep{Pearson1977} returns a p-value of $\approx$0.3, demonstrating that the noise probability density distribution is normal with a high probability (see the example in Fig.~\ref{fig:histogram}).

\subsection{Limiting line flux}

\begin{figure}[htbp]
\begin{center}
\includegraphics[width=\columnwidth]{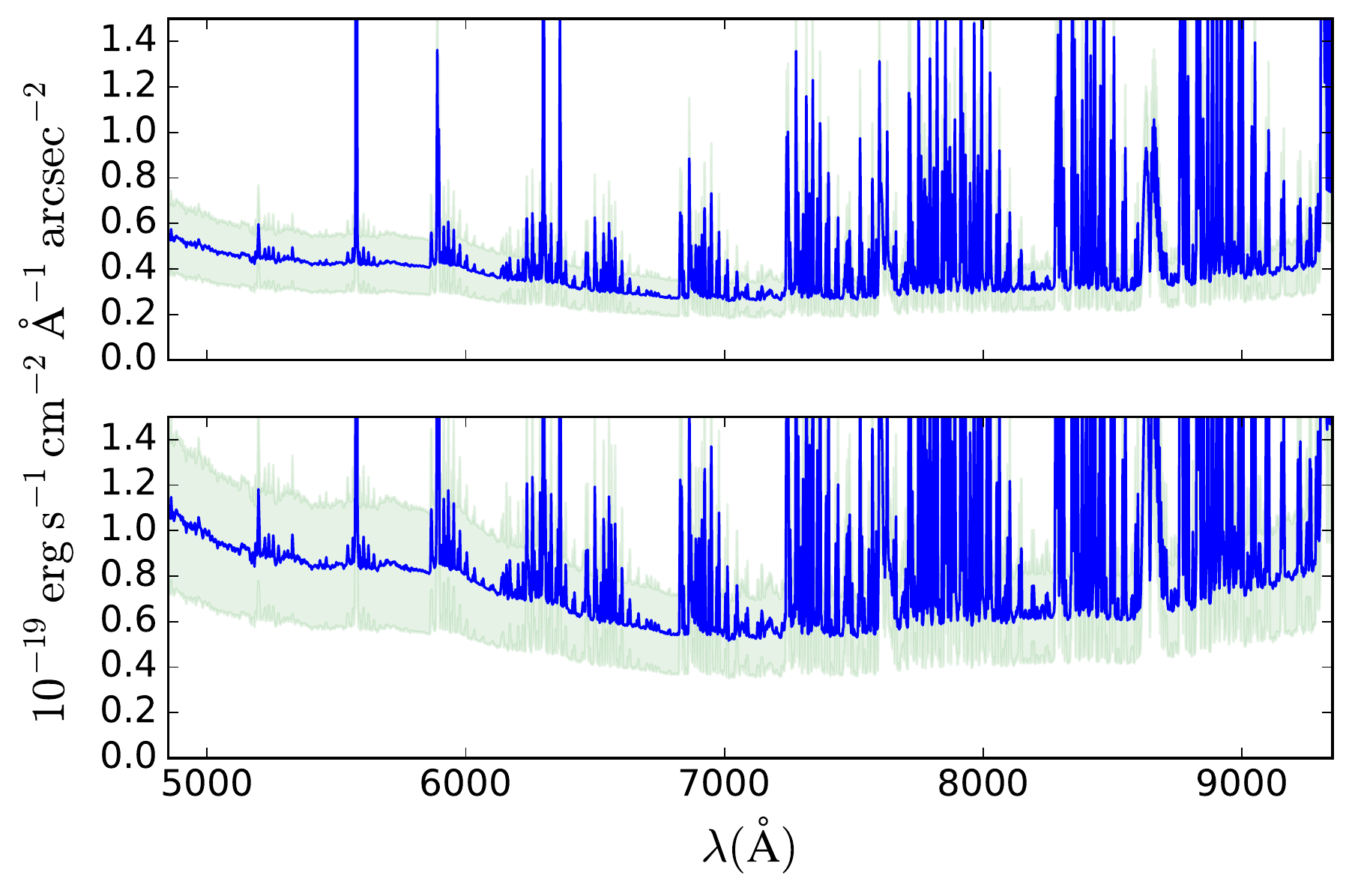}
\caption{$1\sigma$ surface brightness limit for the \mosaic\ (bottom) and \udft\ (top) datacubes computed for an aperture of $1\arcsec \times 1\arcsec$. The blue curve displays the average value and the green area the rms over the field of view.}
\label{fig:SBlimflux}
\end{center}
\end{figure}

From the noise properties one can derive the limiting line flux. 
We start to evaluate the $1\sigma$ emission line surface brightness limit by computing the (sigma-clipped) mean and standard deviation of the propagated variance over the complete field of view for the \udft\ and \mosaic\ datacubes. The resulting emission line surface brightness limit is shown in Fig.~\ref{fig:SBlimflux}. A $1\sigma$ emission line sensitivity  of $2.8$ and $5.5\,10^{-20} \ergs\,\mathrm{arcsec}^{-2}$ for an aperture of $1\arcsec \times 1\arcsec$ is reached in the 7000-8500 \AA\ range for the \udft\ and \mosaic\ datacubes, respectively.

Note that the measured value in the \udft\ (2.8) is slightly better than what we would have predicted from the \mosaic\ value (3.2), taking into account the $\sqrt{3}$ factor predicted by the difference in integration time. It shows that the observational strategy used for the \udft\ (see section~\ref{sect:obs}) is effective in further reducing the systematics which are still present in the \mosaic\ datacube.

This result compares advantageously with the early HDFS observations \citep{HDFS} which reached  a $1\sigma$ emission line surface brightness limit of $4.5\,10^{-20} \ergs \mathrm{arcsec}^{-2}$ in the same aperture. 
The 1.6 better sensitivity\footnote{This factor is probably a lower limit given that the noise analysis performed on the HDFS datacube did not fully take into account the correlated noise} achieved with the \udft\ datacube  is the result of the extensive work performed on observational strategy and data reduction since 2014. While the performance of the first release of the HDFS datacube was dominated by systematics, we have pushed the UDF datacubes to another level of quality and sensitivity. 

\begin{figure}[htbp]
\begin{center}
\includegraphics[width=\columnwidth]{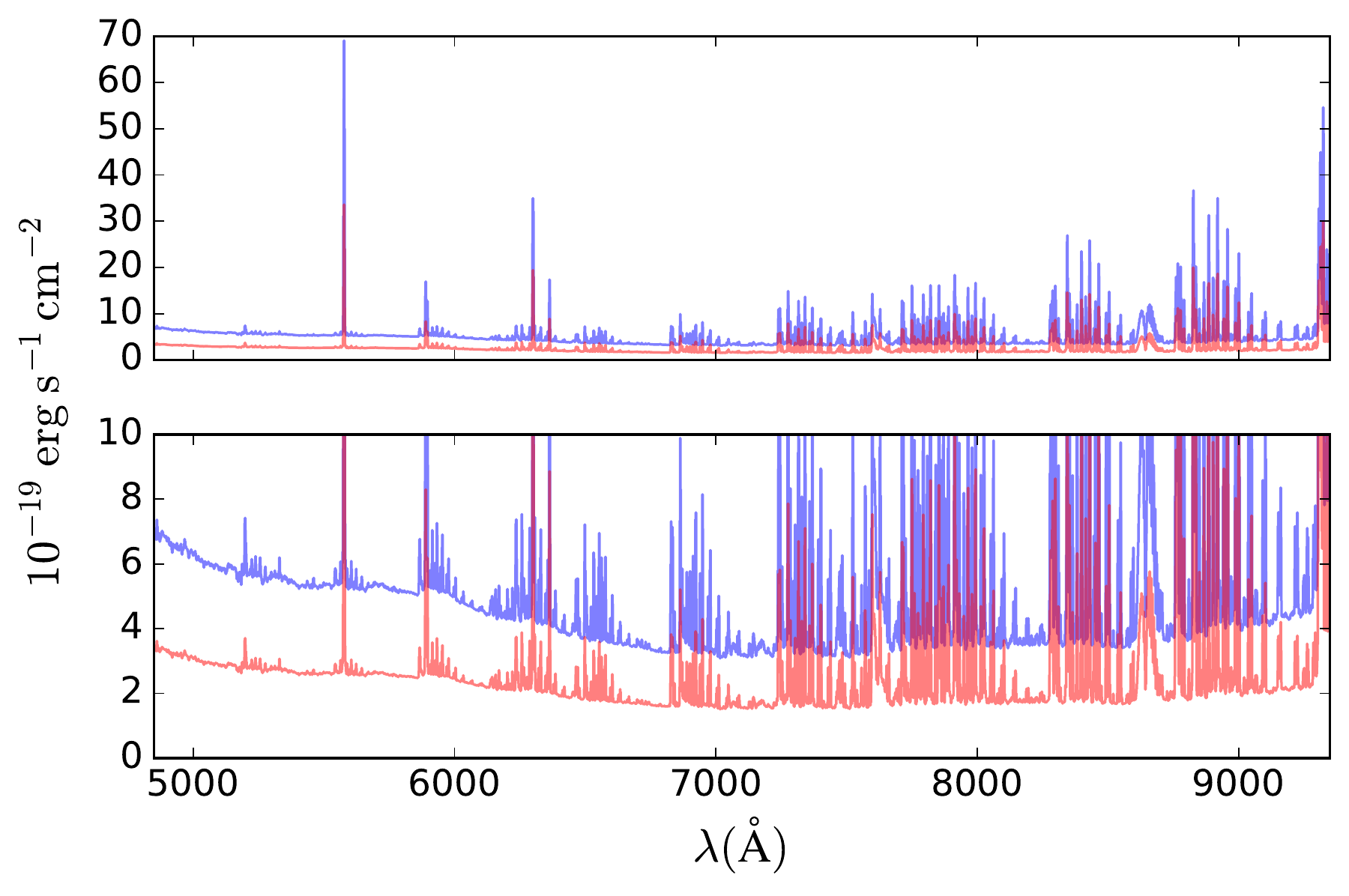}
\caption{$3\sigma$ emission line flux detection limit 
for point-like sources for the \mosaic\ at 10 hours integration time (in blue) and \udft\ at 31 hours integration time (in red) datacubes. 
The full scale sky lines dominated limiting flux is shown in the top panel, while values outside bright sky lines are shown in the bottom panel.}
\label{fig:limitingflux}
\end{center}
\end{figure}

We now derive the line flux detection limit for a point-like source, using weighted FSF extraction and summation over three spectral channels (i.e., 3.75\AA). This value is of course dependent on the integration time (see the exposure map in Fig.~\ref{fig:expmap}). 
We give the $3\sigma$ limiting line flux  in Fig.~\ref{fig:limitingflux} for the corresponding median integration times of the \mosaic\ and \udft\ datacubes. The corresponding detection limits are  $1.5\, 10^{-19}\ergsline$ and $3.1\, 10^{-19}\ergsline$ in the region around 7000 \AA\ between OH sky lines, for the \udft\ and \mosaic\ datacubes, respectively.

\section{Source detection and extraction}
\label{sect:detection}

Exploration of the  \mosaic\ and \udft\ datacubes starts by finding sources, extracting their spatial and spectral information (e.g., subimages and spectra) and measuring their redshifts.  The last step is discussed in paper II \citep{Inami2017}. Here we discuss the first steps using two techniques: optimal source extraction with an HST prior, and blind detection of emission line objects.

\subsection{HST-Prior Extraction}
\label{HSTPRIOR}

As an input to our HST-prior extraction, we use the locations of objects in the \cite{Rafelski2015} source catalog.  This catalog provides precise astrometry, photometry and photometric redshifts for 9927 sources covering the entire UDF region.

Given the MUSE spatial resolution, 0\farcs7 versus 0\farcs1 for HST, our data are unfortunately impacted by source confusion.  Thus, from the inital catalog, we compile a new catalog of 6288 sources, created by merging all \cite{Rafelski2015} sources which have a separation less than 0\farcs6.  For these merged systems, we compute a new source location based on the F775W-light-weighted centroid of all objects that make up the new merged source.

We then proceed to source extraction. Using the \cite{Rafelski2015} segmentation map, we extract each source from the MUSE data in a region defined by its original segmentation area convolved with a Gaussian of 0\farcs6 FWHM to take into account the MUSE resolution. We generate a series of 1D spectra from each extraction region, using several different weighting schemes: (a) a uniformly weighted, direct summation over the full segmentation area, (b) an optimally weighted sum using the reconstructed MUSE white-light image as the weight, and (c) an optimally weighted sum using the estimated FSF at the source location\footnote{In the case of overlapping fields, the FSF is computed as the average of all fields at the source location, weighted by the exposure map.}. 
We also compute a second set of three spectra, using the same weighting schemes, after subtracting a background spectrum from the data.  This spectrum is computed as the average over the empty region free of sources surrounding the object, using the convolved segmentation image as a guide. 

\begin{figure}[htbp]
\begin{center}
\includegraphics[width=\columnwidth]{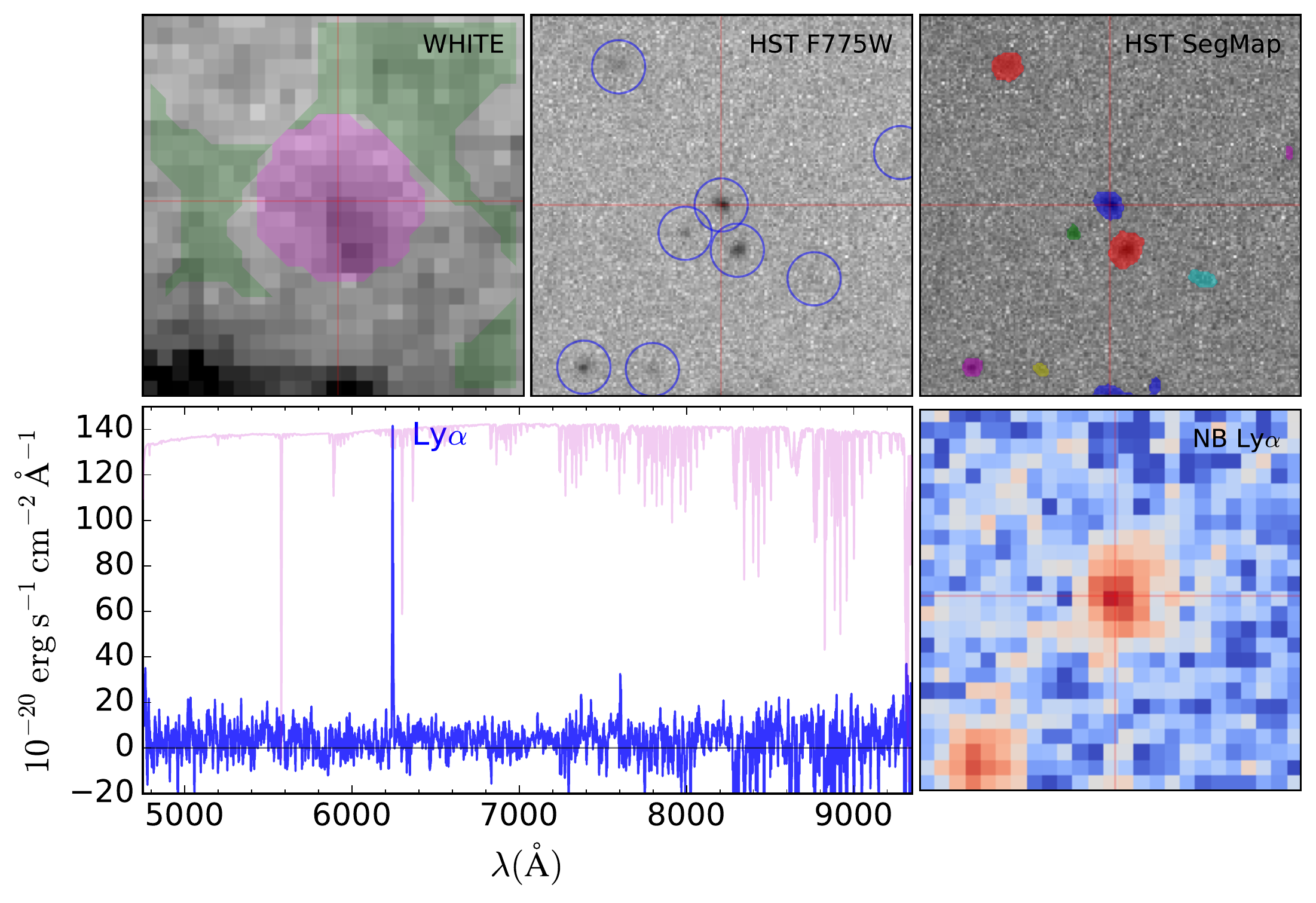}
\caption{Source ID 6698 from the \udft\ data cube. On top, from left to right, one can see the MUSE reconstructed white light image, the HST image in F775W and the HST Rafelski segmentation map. Image size is 5\arcsec\ and the source center is indicated by a red crosshair.  The blue circles mark the sources identified in the Rafelski catalog.  The central Rafelski source ID is 4451 and its F775W AB magnitude is 27.92 $\pm$ 0.04.
The source and background masks are overlaid on the MUSE white light image  in  magenta and green colors, respectively.
Bottom left: PSF weighted extracted source spectrum over the whole wavelength range (box-filtered with a window of 5 pixels). The noise standard deviation is shown in magenta (mirrored with respect to the source spectra). Bottom right: \lya\ Narrow-Band image.}
\label{fig:hstprior}
\end{center}
\end{figure}

The optimal extraction is based on the \cite{Horne1986} algorithm.
\begin{equation}
f(\lambda) = \frac{\sum_x M_x W_{x,\lambda} (D_{x,\lambda} - S_\lambda) / V_{x,\lambda}}{\sum_x M_x W^2_{x,\lambda} / V_{x,\lambda}} 
\end{equation}
\begin{equation}
v_f(\lambda) = \frac{\sum_x M_x W_{x,\lambda} }{\sum_x M_x W_{x,\lambda}^2 / V_{x,\lambda}}
\end{equation}
where $f(\lambda)$ is the optimal flux and $v_f(\lambda)$ its variance, $D$, $S$ and $V$ the data, sky and variance datacubes, $M$ the segmentation mask and $W$ the weight which is either the white-light image or the FSF.
Depending on the object, one of these weighting schemes provides a higher S/N than the others. In general we use the background-subtracted white-light weighted spectra for bright and extended objects ($\mathrm{AB} < 26$ and $\mathrm{FWHM} > 0.5 \times \mathrm{FSF}$)  and background-subtracted FSF weighted spectra for other small and/or faint objects. An example of source extraction is shown in Fig.~\ref{fig:hstprior}.

Due to the convolution, the segmentation map of one source can overlap with other neighboring sources, creating some blending effects in the extracted spectrum.  In a number of cases, as shown in paper II \citep{Inami2017}, the source can be {\em deblended} using the reconstructed narrow-band location when an emission line is present. One such case can be seen in Fig.~\ref{fig:hstprior}. In that figure, the three central HST sources cannot be resolved in the MUSE white light image and thus were originally merged into one source in the extraction process. However, the reconstructed narrow band image shows that the $z = 4.1$ \lya\ emission\footnote{The \lya\ line was identified from its asymmetric profile and fainter continuum on the bluer side of the line} can be clearly attributed to a unique HST object. Note that this galaxy forms a pair with another \lya\ emitter (ID 412) at the same redshift located 3\farcs5 SE with respect to the source center.

The extraction process is run independently for the \mosaic\ and \udft\ datacubes, using the same input catalog in each case to ensure that objects which are in both datacubes receive the same ID. 

\subsection{Blind Detection with ORIGIN}
\label{ORIGIN}

The HDFS study \citep{HDFS} has demonstrated MUSE's ability to detect emission line galaxies without an HST counterpart, so we should not have to rely only on HST-prior source detection when searching for high equivalent-width star-forming galaxies in the UDF. Note however, that the HST data set covering the UDF reaches a $5\sigma$ depth of 29.5 in the F775W filter, i.e., one magnitude deeper than the HST HDFS observations.  Therefore, we expect to find fewer sources without HST  counterparts in the UDF, though this number is surely greater than zero.  Because of this, it is beneficial to attempt to locate these ``hidden'' galaxies through the use of a blind detection algorithm.

Aside from looking for a specific class of galaxy, there is also a practical motivation for performing a blind search of the MUSE datacubes.  As discussed in paper II, redshift assessment is a difficult task which (as of now) is not fully automated, instead relying in large part on expert judgement.  In that respect, investigating all 9927 objects in the \cite{Rafelski2015} catalog is a tedious undertaking.  However, the task can be alleviated by a blind search, assuming it can efficiently pre-select emission line objects.

Several tools have already been developed to perform blind searches of faint emitters in MUSE datacubes, such as: \textsf{MUSELET}, a \textsf{SExtractor} based method available in \textsf{MPDAF}\footnote{See the MPDAF MUSELET documentation at  http://mpdaf.readthedocs.io/en/latest/muselet.html.} \citep{Piqueras2016}, \textsf{LSDCAT}, a matching filter method \citep{LSDCAT}, \textsf{SELFI}, a Bayesian method \citep{2016A&A...588A.140M} and 
\textsf{CubExtractor} (Cantalupo, in prep.), a three-dimensional automatic extraction software based on connecting-labeling-component algorithm (used, e.g., in \citealt{Borisova2016} and \citealt{Fumagalli2016d}).

Each of these methods has its own pros and cons: some achieved high sensitivity but at the expense of low purity, others are optimized to provide reliable results, that is high purity, but with lower sensitivity.  Given the depth and the field of view of the UDF observations, we expect to find thousands of emission line galaxies which, considering the MUSE spatial resolution, will include a significant fraction of blended sources.  The total size of the datacube (3.3 billion voxels for the mosaic) is not negligible either.  In order to handle these methodological and computational challenges, we have begun to develop a new automated method, called  \ori.

The method is still in development and will be presented in a future paper (Mary et al, in prep), but it is already mature enough to be efficiently used for the UDF blind search. In the following sections we briefly explain how the method works and show the results obtained for our observations. 

\subsubsection{Method}

The basic idea of the algorithm is to follow a matched filter approach, where the filters are spatio-spectral (3D) signatures formed by a set of spectral templates (or profiles) that are spatially extended by the point spread function of the instrument \citep{icassp2013}.  In practice, this approach alone is neither robust nor reliable, because the corresponding test statistic is highly sensitive to  sources different from the ones of interest and to residual artifacts (both referred to as unknown nuisance signals). A standard approach in this situation is to model and estimate the nuisance signals under both hypotheses (${\mathcal{H}}_0:$ line absent; ${\mathcal{H}}_1:$ line present), see for instance \cite{kaydetection,scharf94}. However, the resulting tests are computationally intensive and seem hardly compatible with the datacube size. $\ori$ consequently opts for a two-step strategy, where the nuisance signals are suppressed first (using a standard Principal Component Analysis, hereafter PCA) and the lines are detected  in the PCA residuals.  The resulting test statistics are used to assign a probability  to each predetected line. For each line that is flagged as significant, a narrow band (NB) test is performed in order to check whether the line is also significant in the raw data, that is, before any processing (weighting by the estimated variances, PCA) is performed. This step is required because variance underestimation (especially around sky lines, see \ref{sect:variance}) may create artificial lines when weighting the data. Each line that survives the NB test is estimated (deconvolved), leading to an estimate of the line center (a triplet of two spatial and one spectral coordinates). The lines are then merged into sources, leading to a catalog of sources with estimated lines and various other information.

\paragraph{Suppression of nuisance signals:} 
To be consistent with a likelihood based approach,
the whole datacube is first weighted by the estimated standard deviation of the noise in each voxel (Sec.~\ref{sect:variance}). In order to account for spatially varying statistics (regions with more or less bright and/or extended sources) the cube is segmented spatially into several regions (16 for the \udft\ and 121 for the \mosaic).  For a given region, each std-weighted data pixel $\bp$ (a vector whose length is the number of spectral channels) is modeled as a continuum $\bc$ plus a residual $\br$: $\bp=\bc+\br$. The continuum is assumed to belong to a low dimensional subspace, which is obtained by a PCA of all pixels of the considered region. The number of eigenvectors spanning this subspace is computed adaptively for each region. If $\bV_z$ denotes the matrix of the retained (orthonormal) eigenvectors, the residual is estimated as $\widehat{\br}=\bp-\widehat{\bc}=\bp - \bV \bV_z^\top \bp$.
This analysis produces a cube of residuals and, as a side product, a cube of continuum spectra.

\paragraph{Line search:}
For all angular and spectral positions $(\alpha,\delta,\lambda)$ in the residual datacube, the line search considers subcubes of the size of the considered target signatures (typically $13$px $\times \;13$px $\times \;20$ spectral channels, representing $2\farcs6\times2\farcs6\times 25{{\AA}} $) and makes, for each subcube $\bs$ centerd at location $(\alpha_s,\delta_s,\lambda_s)$, a test for the two hypotheses:
\begin{equation*}
\begin{cases}
{\mathcal{H}}_0: \bs = \bn \quad \textrm{(Noise\;only)}, \\
{\mathcal{H}}_1: \bs = \alpha \bSigma^{-\frac{1}{2}} \bd + \bn \quad  \textrm{(Line\;centerd\;at\;$(x_s,y_s,\lambda_s)$\;plus\;noise)},
\label{mod}
\end{cases}
\end{equation*}
where $\bn\sim \mathcal{N}(${\bf{0}}$, ${\bf{I}}$)$ is the noise -- assumed to be a zero-mean Gaussian with an Identity covariance matrix, $\bSigma$ denotes the noise covariance matrix of the data before weighting (assumed to be diagonal in absence of information on noise correlations), $\alpha>0$ is the unknown amplitude
of the emission line and $\bd$ is a spatio-spectral profile weighted by the local values of the noise standard deviation (the weights change for each tested voxel $x_s,y_s,\lambda_s$). The profile  $\bd$ is unknown but assumed to belong to a dictionnary of $12$ spectral profiles (say, $\bd_i,i=1,\hdots,12$) of various widths (from $3$ to $16\,{\AA}$) convolved by the local (wavelength dependent) FSF (a Moffat function). A Generalized Likelihood Ratio (GLR) approach leads to a test statistic $T(\bs)$ in the form of a weighted correlation:
$$T(\bs)=\max_i\frac{\bs^\top \bSigma^{-1} \bd_i}{||\bSigma^{-\frac{1}{2}} \bd_i|| },$$
for which the numerator and denominator can be efficiently computed  using fast convolutions.

\paragraph{$P$-values:}
For the correlations, the $P$-value associated to an observed
 correlation  $t$ is : $p_T :=$ Pr $(T>t\;|\; {\mathcal{}H}_0)$. A $P$-value measures how unlikely a test statistic is under the null hypothesis. The distribution of $T$ under the null hypothesis is estimated from the data in each region and
 $P$-values $p_T$ are computed for each voxel position. All voxels with a $P$-value below a threshold (set after some trial to $10^{-7}$ for the UDF datacubes, corresponding to a detection limit of $5.2\sigma$ for a Gaussian PDF) are flagged as significant.\\
 In the current version, the algorithm also computes a probability that each spectral channel is not contaminated by residual artifacts (such as spurious residuals from sky lines subtraction), by comparing the number of significant $P$-values in each channel against what should be expected from a uniform distribution of noise.\\
The final probabilities (in the form of $P$-values)  evaluate the probability that the line is significant at each voxel position conditionally to the fact that the considered voxel does not belong to  a channel contaminated by artifacts. The  $P$-values less than a threshold (set to $10^{-7}$) survive this step.\\
Thresholding the $P$-values leads to clusters of significant $P$-values, because the signature of a line generally leads to several small $P$-values located in a group of voxels in the vicinity of the line center. To determine a first estimate of the position of the line center, the algorithm retains the smallest $P$-value in each group.

\paragraph{Narrow band tests:}
For each detected line, this step defines a subcube $\bt$ in the raw data centerd on the supposed line location and a control subcube $\bb$ further away in wavelength ($3$ times the spectral length of the profiles that created the detection, say $\bd_k$). A GLR test is then conducted between two hypotheses: under the null hypothesis, both subcubes contain an unknown constant background plus noise, and under the alternative hypothesis $\bt$ also contains the line $\bd_k$ with an unknown amplitude. The test keeps all lines for which the test statistic $\frac{(\bt-\bb)^\top \bd_k}{\sqrt{2}||\bd_k||}$  is larger than a threshold (set to $2$ for the UDF). 

\paragraph{Line estimation:}
The spectral profile and spatial position of each detected line is estimated by spatial deconvolution. The final spectral position of the line is the maximum of the estimated line. Note that while Gaussian profiles
are used for detecting the line, this step allows for the recovery of any line profile, for instance asymmetric or double lines. 

\paragraph{Catalog output:}
The lines are merged into sources by moving over the angular coordinates of the cube containing all detected line centers within a cylinder of diameter equal to the FWHM of the FSF (averaged over the spectral channels) and z axis aligned with the spectral axis. For each object, the algorithm outputs are an ID number, its angular position and the detected lines.  The spectral channel of the line, Gaussian profile that created the initial detection, correlation and spectral channel tests' $P$-values, NB test scores, NB images, deconvolved line profile, estimated flux and FWHM are stored for each line.

\subsubsection{Application to the UDF}
The \ori\ algorithm is implemented as a Python package and was successfully run on the UDF fields using the parameters defined in the previous section. The full computation takes $\approx$1 and $\approx$6 hours of computing time on our 32 multi-core linux workstation for  \udft\ and \mosaic\ datacubes, respectively. The program reported the detection of 355  (\udft) and 1923 (\mosaic) candidate sources. After removing the 49 (\udft) and 672 (\mosaic) false detections\footnote{The false detections are mainly due to residuals left over by  continuum subtraction, splitting of extended bright sources in multiple sources plus a few remaining datacube defects.} identified after visual inspection, we are left with 306 and 1251 potentially {\em real} detections,  corresponding to 86\% and 65\% purity, respectively, for the \udft\ and \mosaic\ fields.

As shown in paper II \citep{Inami2017}, not all detections will eventually turn into a redshift. Generally, the detected sources without redshift have a S/N that is too low to identify the emission and/or absorption lines, but the vast majority at least have an HST counterpart, validating their detection status.

A comparative analysis between the \ori-detected and the HST-prior extracted sources  is presented in paper II. This comparison has been fruitful in finding the remaining problems with \ori\ which impact its sensitivity and/or its purity, which will result in an improved version in the near future. However, despite its current limitations, \ori\ is able to detect a large number of sources, especially high-redshift, faint \lya\ emitters. 

One such example is given in Fig~\ref{fig:orig6524}.  The source is detected at high significance by \ori\ ($P < 10^{-9}$) in the \mosaic\ datacube, as can be seen in the MAXMAP image.  This image is a flattened image of the correlation datacube, displaying the maximum of the correlation over wavelengths. The typical asymmetric \lya\ line profile is very clear, leading to a redshift of 6.24 for this object. Although the source was not identified in the \cite{Rafelski2015} catalog, a faint counterpart is present in the HST F850LP broadband image. The corresponding measured magnitude is AB 29.48 $\pm$ 0.18 (see section \ref{sect:NoiseChisel}). 

The second object (Fig~\ref{fig:orig6326}) is in the \udft\ field. It is also unambiguously detected by \ori\ ($P < 10^{-9}$). The line shape, although less asymmetric than for the previous case, the absence of other emission lines and the undetected continuum, qualifies the galaxy as a \lya\ emitter at z=5.91, but this time one cannot see any HST counterpart. The derived lower limit magnitude is AB 30.7 in the corresponding F850LP broadband filter. In total \ori\ detected 160 sources which were missed in the Rafelski catalog, including  72 which have no HST counterpart (see next section).

\begin{figure}[htbp]
\begin{center}
\includegraphics[width=\columnwidth]{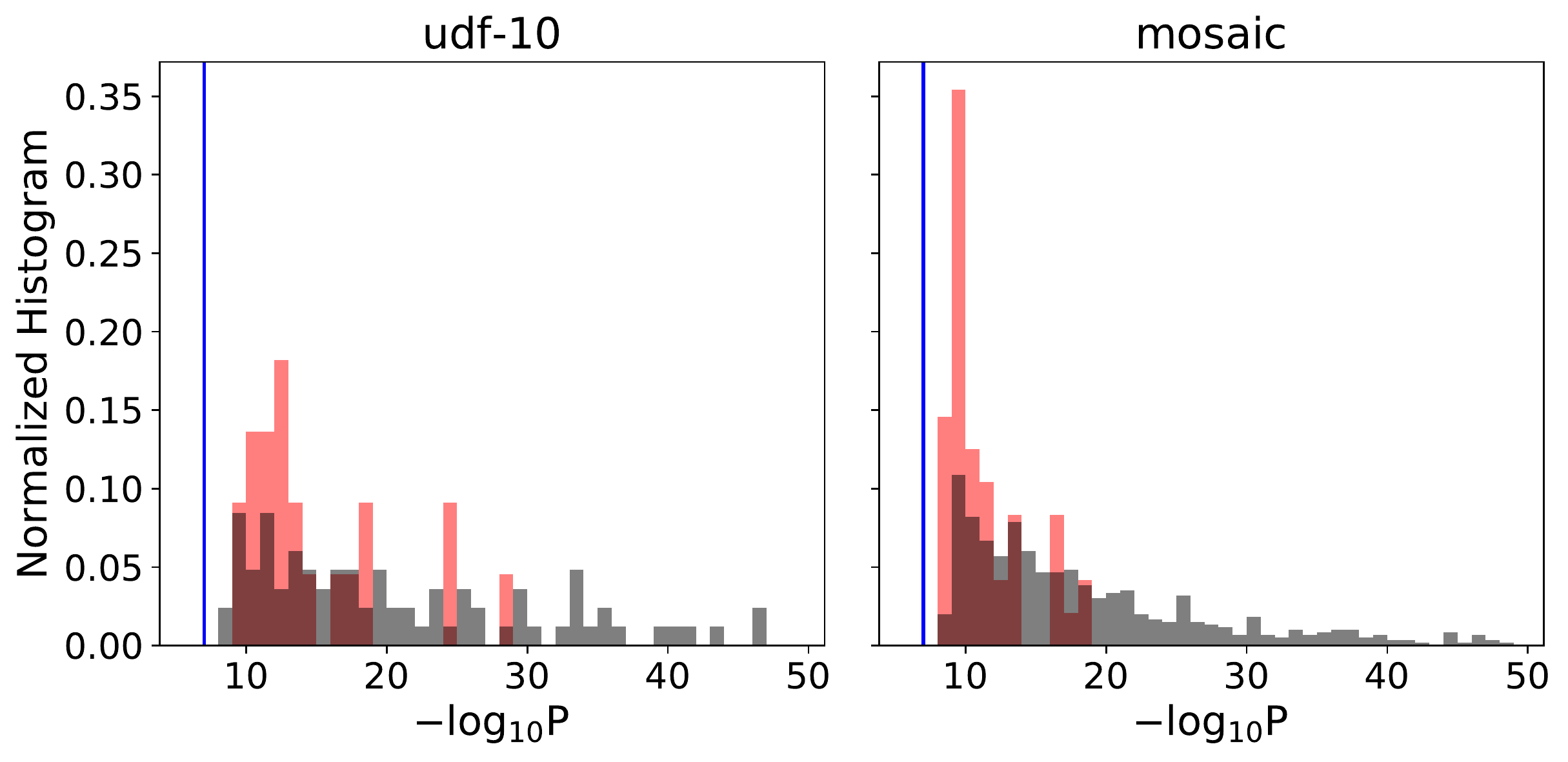}
\caption{Normalized histograms of the P-values of the \ori\ sources with (in gray, restricted to \lya\ emitters) and without (in red) HST counterpart. The blue line displays the threshold P-value ($10^{-7 }$).}
\label{fig:glr}
\end{center}
\end{figure}

\begin{figure*}[htbp]
\begin{center}
\subfloat[Source ID 6524]{
\includegraphics[clip,width=\textwidth]{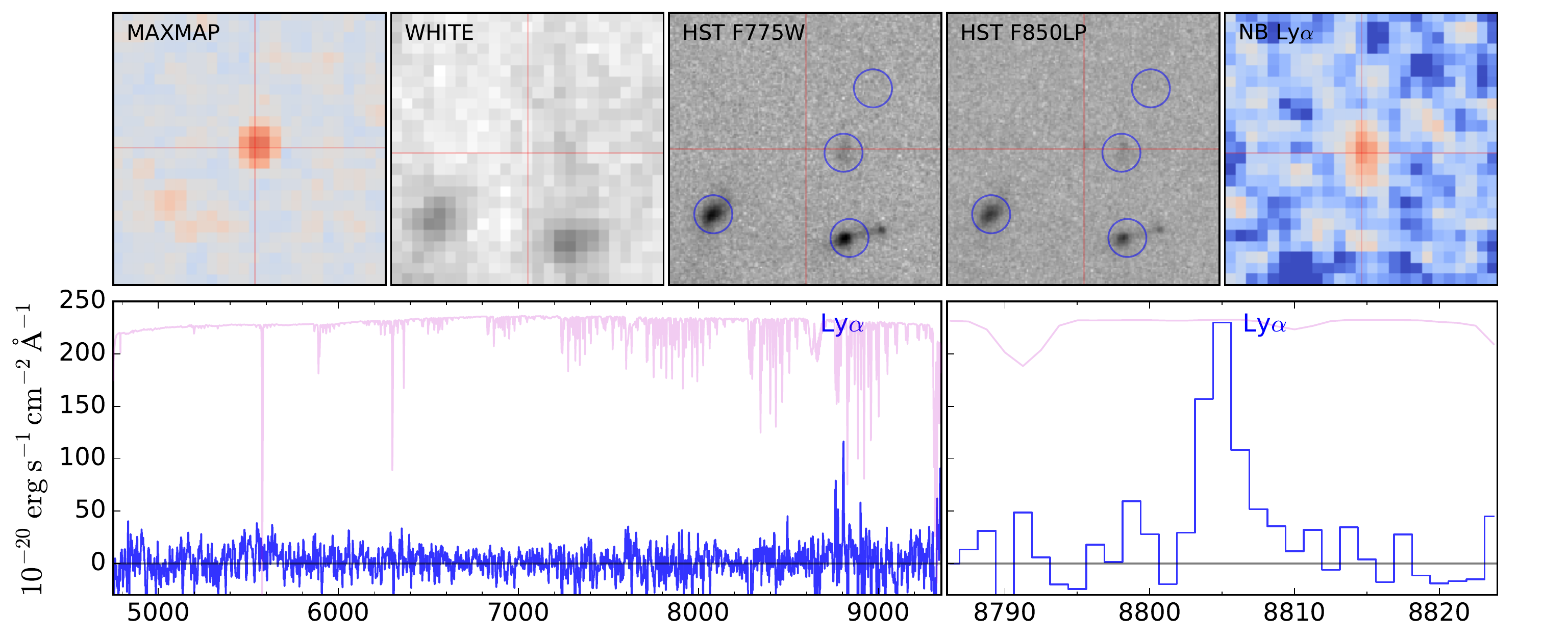}
\label{fig:orig6524}
}
\newline
\subfloat[Source ID 6326]{
\includegraphics[clip,width=\textwidth]{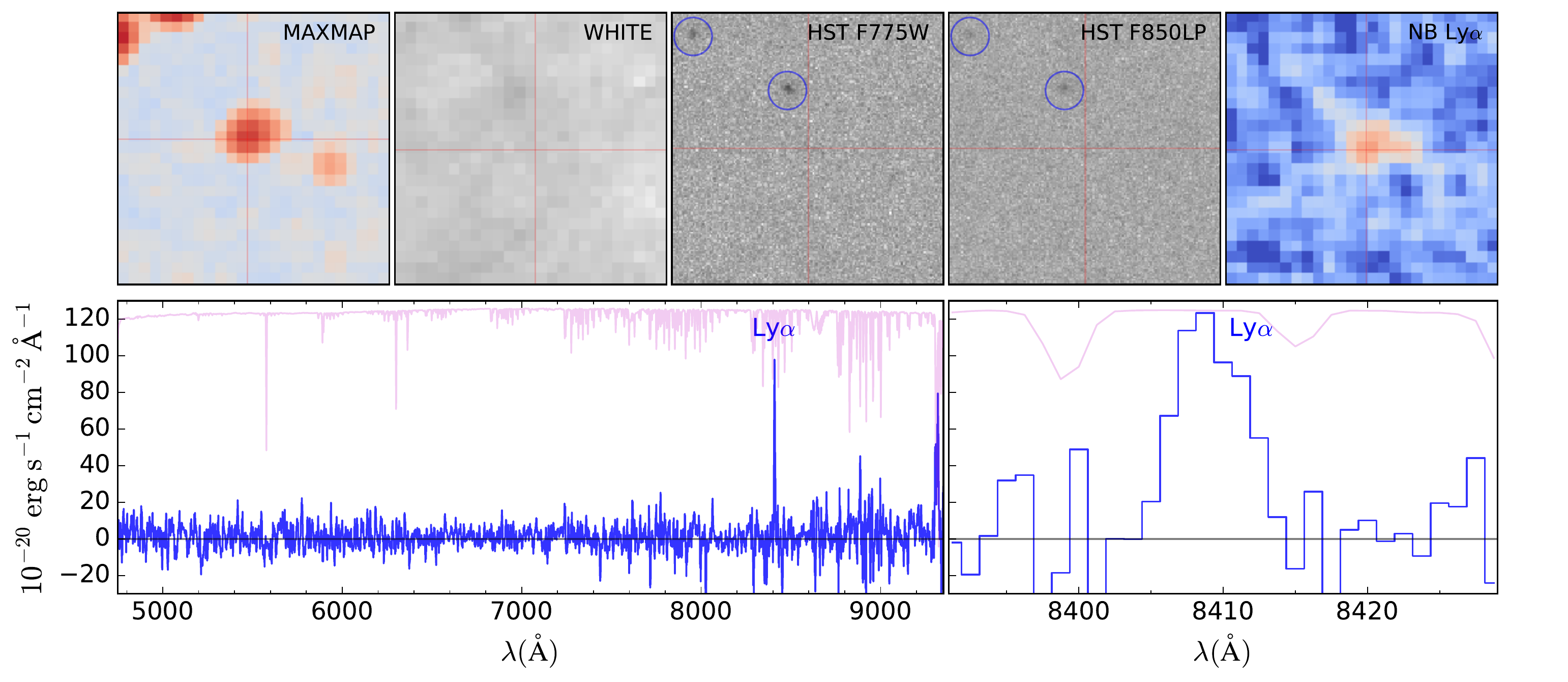}
\label{fig:orig6326}
}
\caption{On top, from left to right, one can see the \ori\ MAXMAP image, the MUSE reconstructed white light image, the HST images in the F775W and F850LP filters and the  \lya\ Narrow-Band image. Image size is 5\arcsec\ and the source center is indicated by a red crosshair.  The blue circles mark the sources identified in the Rafelski catalog. Bottom: source spectrum over the whole wavelength range (box-filtered with a window of 5 pixels) and zoomed (unfiltered) around the \lya\ line. The noise standard deviation is shown in magenta (mirrored with respect to the source spectra).}

\end{center}
\end{figure*}

We investigate how reliable is the detection of these 72 new sources by comparing their P-values  with the corresponding values of the \ori\ detections (restricted to \lya\ emitters) and successfully  matched with an HST source. The histograms of the P-values  for the two populations are given in Fig.~\ref{fig:glr}. As expected, the sources with low P-values ($<10^{-29 }$, $<10^{-18}$ in  \udft\ and \mosaic\ respectively ) are all detected in HST. However, except for these bright emitters, the P-values of the \textit{HST-undetected} sources are not very different from the general population. This is especially true for  \udft\  which goes deeper than \mosaic. At similar P-values, the sources detected by \ori\ with HST counterpart were unambiguously identified (see paper II for the detailed evaluation) giving confidence that most of the \textit{HST-undetected} sources found by \ori\ are real.

\subsection{HST photometry of newly detected sources}
\label{sect:NoiseChisel}

We performed a simple aperture photometric analysis by computing  HST AB magnitudes in a 0\farcs4 diameter centered at the source location for all HST broadband images\footnote{Note that these fixed aperture magnitudes can be different from those given in paper II which are based on the \noisechisel\ segmentation maps}. The magnitudes were compared to the $5 \sigma$ detection limit  of the corresponding HST filter (see column $\mathrm{AB_{5 \sigma}}$ in Table \ref{tab:nohstmag}). A source is defined as \textit{HST-detected} when it is brighter than the $5 \sigma$  detection limit in at least one of the HST filters. 
Note that for the sources which fall outside the region with the deepest WFC3 IR data the corresponding shallower limiting depth was used. 

\begin{figure}[htbp]
\begin{center}
\includegraphics[width=\columnwidth]{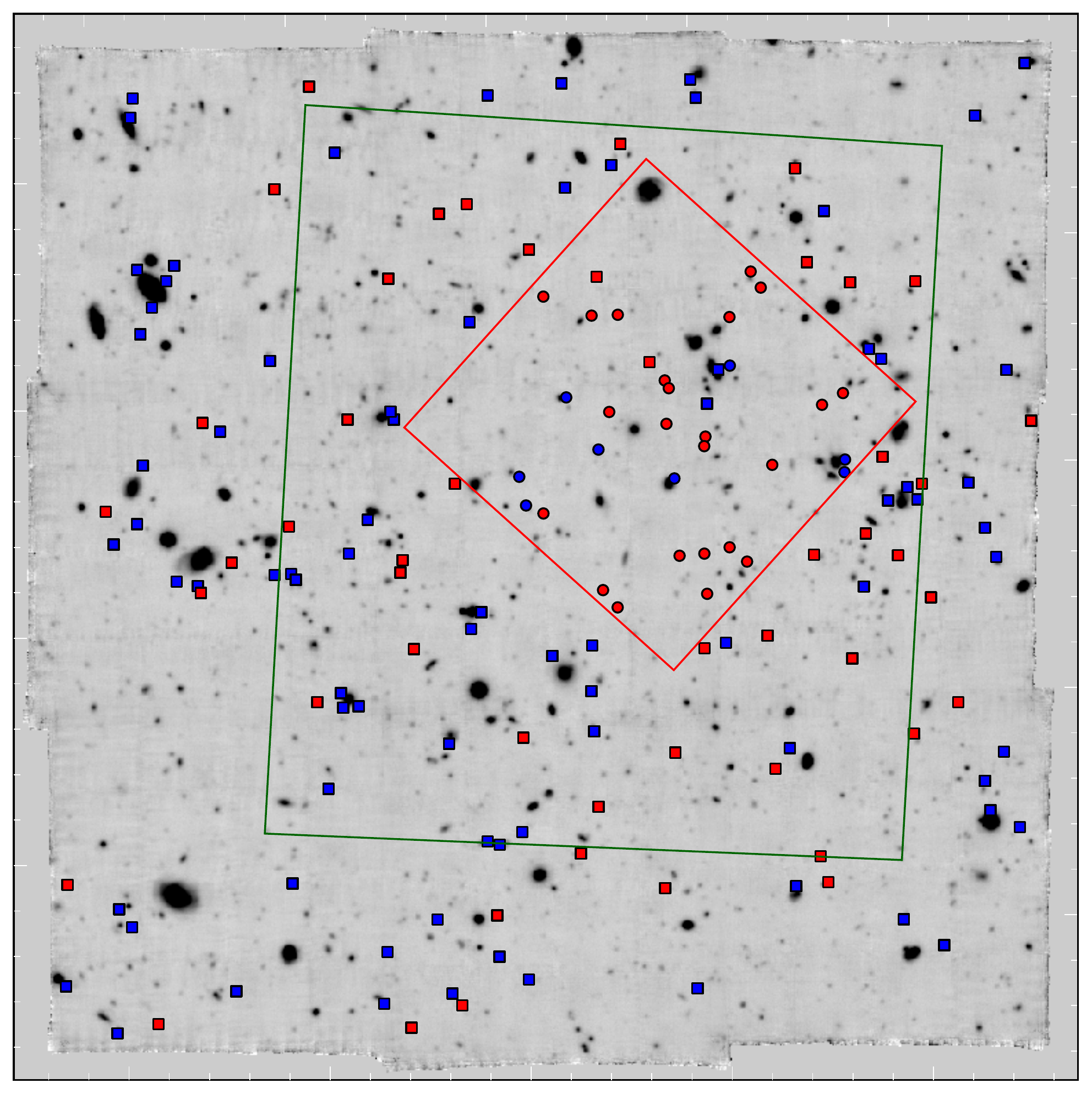}
\caption{Location of the new sources detected by \ori\ overlaid on the \mosaic\ white-light image.  \textit{HST-detected} objects, i.e., brighter than the detection depth in at least one HST filter, are shown in blue, while the \textit{HST-undetected} ones are displayed in red. The \udft\ and \mosaic\ sources are marked with a circle and a square symbol, respectively.
The green rectangle indicates the XDF/HUDF09/HUDF12 region containing the deepest near-IR observations from the HST WFC3/IR camera. The red square show the \udft\ field location.
The north is located 42\degr\ clockwise from the vertical axis. }
\label{fig:nohstloc}
\end{center}
\end{figure}

The location of all sources without \cite{Rafelski2015} catalog entries are shown in Fig.\ref{fig:nohstloc}. Among these 160 sources, 72 where considered as \textit{HST-undetected}, i.e., with all computed magnitude larger than the detection limit. While the majority of these objects (54) are located within the region with the deepest WFC3 IR data, a small fraction (18) are found outside this region.  Although all of these objects without HST counterpart fall below the detection limit of Rafelski, we derive a rough estimate of their average magnitude by computing the mean AB magnitude and its standard deviation for the entire sample of 54 sources present in the area of the HST deepest IR images (Table \ref{tab:nohstmag}). 
A detailed analysis of the properties of these sources is deferred to another paper (Maseda et al, in prep).

\begin{table}
\caption{Mean HST AB magnitude ($\mathrm{\overline{AB}}$) of the 54 sources without HST counterpart in the deepest UDF region (displayed as a green rectangle in Fig.~\ref{fig:nohstloc}).
The reference AB $\mathrm{5 \sigma}$ depth ($\mathrm{AB_{5 \sigma}}$) from Table 1 of  \cite{Rafelski2015} is shown.}
\label{tab:nohstmag}
\begin{center}
\begin{tabular}{ccc}
\toprule
Filter & $\mathrm{\overline{AB}}$  & $\mathrm{AB_{5 \sigma}}$ \\
\midrule
F606W & 31.8 $\pm$ 1.1 & 29.6 \\
F775W & 31.3 $\pm$ 1.3 & 29.5 \\
F850LP & 31.0 $\pm$ 1.4 & 28.9 \\
F105W & 31.3 $\pm$ 0.9 & 30.1 \\
F125W & 31.1 $\pm$ 0.6 & 29.7 \\
\bottomrule
\end{tabular}
\end{center}
\end{table}

We inspect the 88 \textit{HST-detected} objects discussed above to understand why they were missing in that catalog. We found three main reasons:  1) distant deblending, where the object is clearly detected but parametric fitting had associated it with a distant neighbor, see Figure 17 in \citet{NoiseChisel}; 2) nearby deblending, where the object was too close to a bright object  to be identified as a separate object; and 3) Manual removal based on S/N after running \textsf{SExtractor}, to correct for low purity. 
These three classes constituted 8\%, 73\%, and 15\% of the missed objects.

To perform  optimal source extraction as presented in section \ref{HSTPRIOR}, we update the Rafelski segmentation map with the segments corresponding to the new detected object.
\citet{Rafelski2015} had already used multiple
\textsf{SExtractor} \citep{Bertin+1996} runs to generate their segmentation map. Hence for image segmentation and broadband measurements of these objects, we
adopted \noisechisel\ \citep{NoiseChisel}. \noisechisel\ is non-parametric and much less sensitive to the diffuse flux of the neighboring objects. Therefore it is ideally suited to complement the \citet{Rafelski2015}
catalog.

\noisechisel\ was configured to ``grow'' the detected ``clumps'' into the
diffuse regions surrounding them when there are no other clumps
(resolved structure) over the detection area (see Figure 10 of \citealt{NoiseChisel}). 
The final segmentation map for each object was selected as the one which gives the largest detection area among all filters. 
Checking the
correspondence between magnitudes derived with this configuration and with \citet{Rafelski2015}, 
we found the expected agreement in
derived magnitudes: 
that is, in the AB magnitude interval
$\rafcompbincenter\pm\rafcompbinpm$, the
$\rafcompscmultip\sigma$ iterative clipped rms (terminated when the relative
change in rms goes below $\rafcompsctolerance$) was $\rafcompdiffstd$
in the \rafcompfilt{} filter. As a comparison, the R15 catalog has rms
of $\rafcomporigstd$ with the same magnitude interval, filter and
method.

\begin{figure*}[htbp]
\begin{center}
\includegraphics[width=\textwidth]{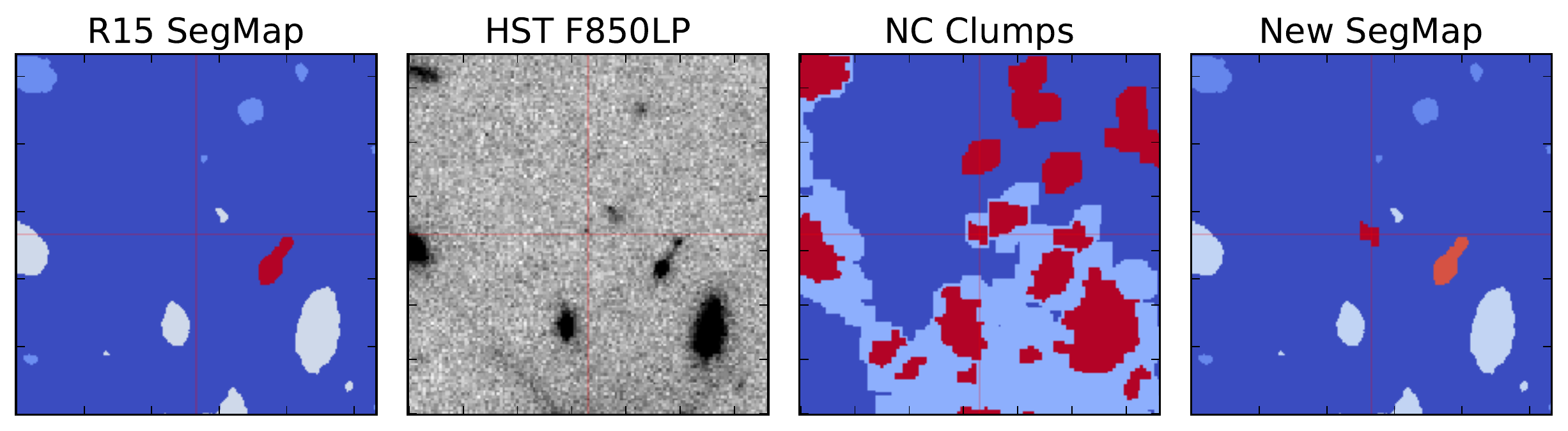}
\caption{Complementing the \citet{Rafelski2015} segmentation map with 
\noisechisel\ on source ID 6524 (see also Fig.~\ref{fig:orig6524}). Note that images are displayed 
in the original HST grid (rotated by -42\degr\ compared to Fig.~\ref{fig:orig6524}). 
Image size is 8\arcsec and the target is in the center (shown by the red crosshair). From 
left to right: The \citet{Rafelski2015} segmentation map, the input F850LP 
image, \noisechisel\ clumps (red) over diffuse detections (light blue), and 
the final segmentation map, with the central clump of the previous image 
added to the input segmentation map. Note how some red regions in the 
\noisechisel\ clumps image are not surrounded by diffuse flux (light blue). The measured magnitude is 29.49$\pm$0.18 in the F850LP filter. See Section~\ref{sect:NoiseChisel} for more details.}
\label{fig:noisechisel}
\end{center}
\end{figure*}

\noisechisel\ detected the previously mentioned objects, along
with another $\rafprobnc\%$ of the initial sample. For the remaining
objects, an aperture of diameter 0\farcs5 was placed on the
position reported by \ori. Each object's footprint was
randomly placed in $\mkcatupnum$ non-detected regions and the
$\mkcatupnsigma\sigma$ width of the final distribution was defined as an
upper limit on the magnitude. In the case of the WFC3/IR images that contain
the wide HUDF and deep XDF/IR depths, this was done on the depth the
object was positioned in, not the full UDF area. When the object's
magnitude was below the upper limit magnitude in a filter, the latter
was used in the catalog.
An example of a \noisechisel\ detection performed on one of the sources without a \citet{Rafelski2015} catalog entry (ID 6524 Fig.~\ref{fig:orig6524}) is presented in Fig.~\ref{fig:noisechisel}.

\section{Summary and conclusion}

In this first paper of the series, we have presented the MUSE observational campaign of the Hubble Ultra Deep Field for a total of 137 hours of VLT time,  performed in 2014 and 2015 over eight runs of our Guaranteed Time Observing. A contiguous area of 9.92 \amind\ was observed with a \mosaic\ of nine fields. It covers almost the entire UDF region at a median depth of 9.6 hours. A single field (\udft) of 1.15 \amind\ located within the XDF region, was also observed at additional depth. When combined with the \mosaic\ fields, it reaches a median depth of 30.8 hours.

The reduction of this large data set was performed using an advanced scheme to better remove the systematics and improve the overall quality of the produced datacubes. An enhanced self-calibration process, a better masking of instrument artefacts and the use of the PCA ZAP \citep{ZAP} software to remove sky residuals, results in datacubes with improved quality with respect to the previous HDFS MUSE observations and data reduction \citep{HDFS}.

We investigated the astrometry and broadband photometric properties of the datacubes, using the HST deep images as reference. We found an astrometric accuracy of  0\farcs07 rms, i.e., $\frac{1}{10}$ of the spatial resolution, for galaxies brighter than AB 27. We also assessed the broadband photometric performance, still using HST magnitude as reference. Although the achieved photometric accuracy of MUSE datacubes cannot compete with the performance of the UDF HST deep broadband imaging, especially in the redder part of the spectrum dominated by OH lines, we found good agreement with little systematic offset up to magnitude AB 28. The scatter of MUSE magnitudes with respect to HST is 0.4 magnitudes in F606W for the \udft\ data cube at  AB 26.5, and 0.8 magnitudes for the F775W and F814W filters at the same magnitude.

We developed an original method to accurately measure the spatial resolution of the observations through a comparison with the HST broadband images. This method can be used when there is no bright star in the MUSE field. It works in Fourier space and also provides a good estimate of the absolute astrometric  and photometric offsets with respect to HST. Using this new tool, we derived the spatial PSF of the combined datacubes, modeled as a Moffat function with a constant $\beta=2.8$ parameter and a linear decrease of FWHM with wavelength. The achieved spatial resolution (Fig.~\ref{fig:fsffields}) is 0\farcs71 (at 4750\AA) and 0\farcs57 (at 9350\AA) FWHM for both the \mosaic\ and \udft\ fields. There is little dispersion for the \mosaic\ sub-fields, with a measured standard deviation of only 0\farcs02. 

We investigated the noise properties of the two final datacubes. The noise distribution is well represented by a Normal probability density function. The empirical correction accounting for the correlated noise in each individual datacube prior to the combination works well. The final corrected propagated standard deviation is a good representation of the true noise distribution in regions with faint sources (e.g., dominated by the sky noise). 
A $1\sigma$ surface brightness emission line sensitivity  (Fig.~\ref{fig:SBlimflux}) of $2.8$ and $5.5\,10^{-20} \ergs \mathrm{arcsec}^{-2}$ is reached in the red for an aperture of $1\arcsec \times 1\arcsec$ and for the \udft\ and \mosaic\ datacubes, respectively. This is a factor 1.6 better than the sensitivity measured in the first release of the HDFS datacube, demonstrating the progress achieved in the data reduction and observational strategy. 
A 3$\sigma$ point source line detection limit (Fig.~\ref{fig:limitingflux}) of $1.5$ and $3.1\,10^{-19} \ergsline$  is achieved in the red (6500-8500\AA) and between OH sky lines for the \udft\ and \mosaic\ datacubes,  respectively. 

We extracted 6288 and 854 sources from the \mosaic\ and \udft\ datacubes, using the \cite{Rafelski2015} catalog and segmentation map as input for the source locations. For each source we performed optimal extraction, weighted with either the white light image or the FSF at the source location. A large number (40\%) of HST sources are blended at the MUSE spatial resolution, but we show that this blending can often be resolved using reconstructed narrow-band images to locate sources that have detected emission lines.

In parallel we performed a blind search for emission line objects using an algorithm (\ori) developed specically for MUSE datacubes. \ori\ computes test statistics on a matched filtered datacube after a PCA-based continuum removal. The blind search results in 306 and 1251 detections in the \udft\ and \mosaic\ datacubes, respectively. 

A number of these sources (160) were not present in the \cite{Rafelski2015} catalog. Investigation of these new sources show that 55\% of them are bright enough in at least one of the HST band to be detected, but have been missed because of contamination and/or uncorrect \textsf{SExtractor} deblending process.
The remaining 72 sources fall below the detection limit of HST broadband deep images. In the HST region with deep WFC3/IR images, we compute a mean AB magnitude of 31.0 - 31.8 within a 0\farcs4 diameter aperture.
We use \noisechisel, a \textsf{SExtractor} alternative optimized for the detection of diffuse sources, to derive an updated segmentation map for these sources when that was possible.

The redshift measurement and analysis of this unprecedented data set is presented in paper II \citep{Inami2017}. With more than 1300 high-quality redshifts, this survey is the deepest and most comprehensive spectroscopic study of the UDF ever performed. It expands the present spectroscopy data set (173 galaxies accumulated over ten years) by almost an order of magnitude and covers a wide range of galaxies, from nearby objects to $z = 6.6$ high redshift \lya\ emitters, and from bright (magnitude 21) galaxies to the faintest objects (magnitude $>30$) visible in the HST images.

\begin{figure*}[htbp]
\begin{center}
\includegraphics[height=0.95\textheight]{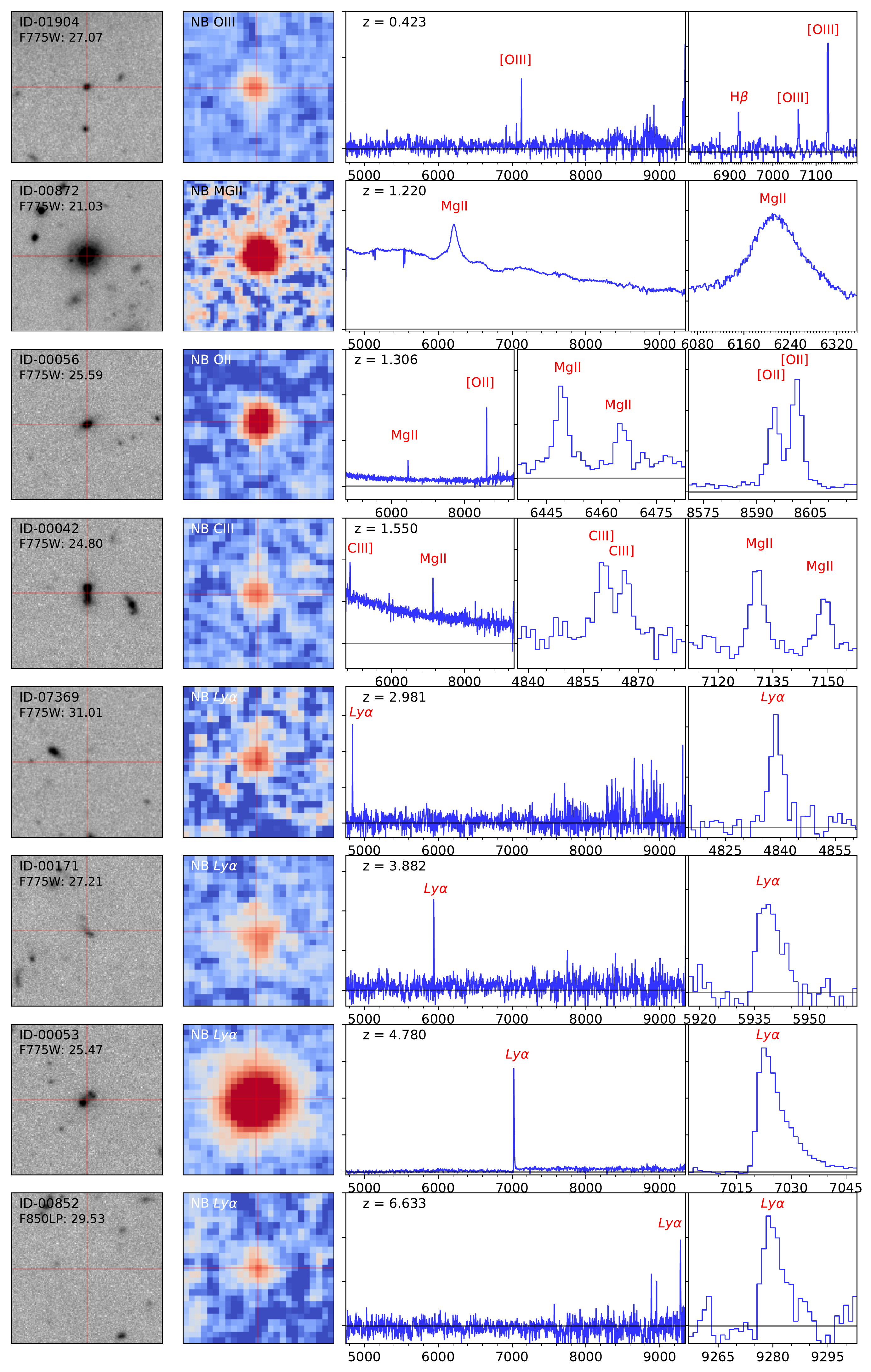}
\caption{Example of sources from the \mosaic\ and \udft\ fields. Each row shows a different object, ordered by redshift. From left to right one can see: the HST broadband image (F775W filter), a MUSE-reconstructed narrow-band image of one of the brightest emission lines, the source spectrum over the full wavelength range and a zoom-in region highlighting some characteristic emission lines. The images have a linear size of 5\arcsec\ and the source center is displayed as a red cross-hair.}
\label{fig:sources}
\end{center}
\end{figure*}

Of course, the survey "performance"  is much more than just the number of faint sources from which we are able to obtain reliable redshifts. The quality of the MUSE data, as shown in Fig.~\ref{fig:sources} for a few representative sources, enables new and detailed studies of the physical properties of the galaxy population and their environments over a large redshift range. In subsequent papers of this series, we will therefore explore the science content of this unique data set.

\begin{acknowledgements}
RB,SC,HI,JBC,MS,MA acknowledges support from the ERC advanced grant 339659-MUSICOS. 
JR, DL acknowledges support from the ERC starting grant CALENDS.
RB,TC,BG,NB,BE,JR acknowledges support from the FOGHAR Project with ANR Grant ANR-13-BS05-0010. JS acknowledges support from the  ERC grant 278594-GasAroundGalaxies. SC and RAM acknowledges support from Swiss National Science Foundation grant PP00P2 163824. Part of this work was granted access to the HPC and visualization resources of the Centre de Calcul Interactif hosted by University Nice Sophia Antipolis. BE acknowledges financial support from ``Programme National de Cosmologie et Galaxies'' (PNCG) of CNRS/INSU, France. JB acknowledges support by Funda{\c c}{\~a}o para a Ci{\^e}ncia e a
Tecnologia (FCT) through national funds (UID/FIS/04434/2013) and Investigador FCT 
contract IF/01654/2014/CP1215/CT0003., and by FEDER through COMPETE2020 (POCI-01-0145-FEDER-007672). 
PMW received support through BMBF Verbundforschung (project MUSE-AO, grant 05A14BAC).
\end{acknowledgements}

\bibliographystyle{aa}
\bibliography{udf-biblio.bib}

\end{document}